%% file: gxpap_time.tex
\documentclass{article}
\usepackage{psfig}
\usepackage{rotating}
\usepackage{epsf}
\usepackage{emulateapj}
\usepackage{apjfonts}

\pssilent

\makeatletter
\let\@internalcite\cite
\def\cite{\def\astroncite##1##2{##1\ ##2}\@internalcite}
\def\citey{\def\astroncite##1##2{##1\ (##2)}\@internalcite}
\def\@citex[#1]#2{\if@filesw\immediate\write\@auxout{\string\citation{#2}}\fi
  \def\@citea{}\@cite{\@for\@citeb:=#2\do
    {\@citea\def\@citea{; }\@ifundefined
       {b@\@citeb}{{\bf ??}\@warning
       {Citation `\@citeb' on page \thepage \space undefined}}%
{\csname b@\@citeb\endcsname}}}{#1}}
\def\@cite#1#2{#1\if@tempswa #2\fi}
\def\@biblabel#1{}
\def\astroncite#1#2{#1\ #2}
\makeatother

\def\errtwo#1#2#3{ $#1^{+ #2}_{- #3}$ }

\def\aproxgt{\mathrel{%
      \rlap{\raise 0.511ex \hbox{$>$}}{\lower 0.511ex \hbox{$\sim$}}}}
\def\aproxlt{\mathrel{%
      \rlap{\raise 0.511ex \hbox{$<$}}{\lower 0.511ex \hbox{$\sim$}}}}

\def\be{\begin{equation}}
\def\ee{\end{equation}}
\def\bea{\begin{eqnarray}}
\def\eea{\end{eqnarray}}

\begin{document}

\slugcomment{To Be Published in The Astrophysical Journal}

\lefthead{Nowak et al.}
\righthead{Low Luminosity States of the Black Hole Candidate
  GX~339--4. II.}

\title{Low Luminosity States of the Black Hole Candidate
  GX~339--4. II. Timing Analysis}

\author{Michael A. Nowak\altaffilmark{1}, J\"orn Wilms\altaffilmark{2},
James B. Dove\altaffilmark{1,3}}

\altaffiltext{1}{JILA, University of Colorado, Campus Box 440, Boulder, CO
     ~80309-0440, USA; \{mnowak, dove\}@rocinante.colorado.edu}
\altaffiltext{2}{Institut f\"ur Astronomie und Astrophysik,
     Abt.~Astronomie,  Waldh\"auser Str. 64, D-72076 T\"ubingen,
     Germany; wilms@astro.uni-tuebingen.de} 
\altaffiltext{3}{also, CASA, University of Colorado, Campus Box 389,
     Boulder, CO ~80309-0389, USA} 

\received{September 1, 1998}
\accepted{September 2, 1998}

\begin{abstract}
  
  Here we present timing analysis of a set of eight Rossi X-ray Timing
  Explorer (RXTE) observations of the black hole candidate GX~339--4 that
  were taken during its hard/low state.  On long time scales, the RXTE All
  Sky Monitor data reveal evidence of a 240 day periodicity, comparable to
  timescales expected from warped, precessing accretion disks.  On short
  timescales all observations save one show evidence of a persistent
  $f_{\rm QPO} \approx 0.3$\,Hz QPO.  The broad band
  ($10^{-3}$--$10^2$\,Hz) power appears to be dominated by two independent
  processes that can be modeled as very broad Lorentzians with $Q \aproxlt
  1$.  The coherence function between soft and hard photon variability
  shows that if these are truly independent processes, then they are
  individually coherent, but they are incoherent with one another.  This is
  evidenced by the fact that the coherence function between the hard and
  soft variability is near unity between $5\times10^{-3}$--10\,Hz but shows
  evidence of a dip at $f \approx 1$\,Hz. This is the region of overlap
  between the broad Lorentzian fits to the PSD.  Similar to Cyg~X--1, the
  coherence also drops dramatically at frequencies $\aproxgt 10$\,Hz.  Also
  similar to Cyg~X--1, the hard photon variability is seen to lag the soft
  photon variability with the lag time increasing with decreasing Fourier
  frequency.  The magnitude of this time lag appears to be positively
  correlated with the flux of GX~339--4.  We discuss all of these
  observations in light of current theoretical models of both black hole
  spectra and temporal variability.

\end{abstract}



\setcounter{footnote}{0}

\section{Introduction}\label{sec:intro}

In a companion paper to this work (\cite{wilms:98b}; hereafter paper I) we
have presented spectral analysis of a series of Advanced Satellite for
Cosmology and Astrophysics (ASCA) and simultaneous radio/Rossi X-ray Timing
Explorer (RXTE) observations of the black hole candidate (BHC) GX~339--4.
This source has exhibited both spectrally soft states (cf.
\cite{grebenev:91a,miyamoto:91a}) and spectrally hard states (cf.
\cite{grebenev:91a,miyamoto:92a,zdziarski:98a}).  Both the ASCA and RXTE
observations presented in paper I showed GX~339-4 to be in spectrally hard
and low luminosity (3--9\,keV flux $\aproxlt 10^{-9}~{\rm
  ergs~cm^{-2}~s^{-1}}$) states.  The eight RXTE observations spanned
roughly a factor of five in terms of observed 3--9\,keV flux.

In this work we shall consider the timing analysis of the eight RXTE
observations.  Timing analysis for GX~339--4 previously has been presented
for the soft, bright `very high state' (\cite{miyamoto:91a}), the soft,
fainter `high state' (\cite{grebenev:91a}), and the hard `low state'
(\cite{grebenev:91a,miyamoto:92a}).  In addition, timing analysis has been
presented for so-called `intermediate states', which are bright but
spectrally harder than the typical `very high' or `high states'
(\cite{mendez:97a}).  In general, the harder states exhibit more temporal
variability than the softer states (cf. \cite{vanderklis:89b}).  Timing
analyses of the GX~339--4 hard state have shown similar results to analyses
of other hard state BHC (\cite{miyamoto:92a}).  Specifically, the GX~339--4
temporal variability has been observed to be similar to that of Cygnus~X--1
(\cite{belloni:90a,belloni:90b,miyamoto:92a,nowak:98a}, and references
therein).

A discussion of Fourier techniques in specific, and timing analysis in
general, has been presented by \citey{vanderklis:89b}.  Here we apply these
Fourier analysis techniques in the same manner as for our RXTE observations
of Cyg X--1 (\cite{nowak:98a}).  Specifically, we used the same techniques
for estimating: deadtime corrections (\cite{zhangw:95a,zhangw:96a}); the
error bars and Poisson noise levels of the Power Spectral Density (PSD)
(\cite{leahy:83a,vanderklis:89b}); the error bars and noise levels for the
coherence function (\cite{bendat,vaughan:97a}); and the error bars and
noise levels for the Fourier frequency-dependent time lag between hard and
soft photon variability (\cite{bendat,nowak:98a}).  A self-contained
discussion of these techniques as regards RXTE timing analysis 
is given by \citey{nowak:98a}.

This paper is organized as follows.  First we consider evidence for long
term periodicities in the RXTE All Sky Monitor (ASM) data in
section~\ref{sec:asm}.  In section~\ref{sec:psd} we consider the Power
Spectral Density (PSD). In section~\ref{sec:lag} we consider both the
coherence function (cf. \cite{vaughan:97a}) and the Fourier
frequency-dependent time lags between hard and soft photon variability.  We
discuss the theoretical implications of these observations in
section~\ref{sec:discuss}.  We then summarize our results in
section~\ref{sec:summ}.

\section{All Sky Monitor Observations of Long `Characteristic Timescales'
  in GX~339$-$4}\label{sec:asm}

\begin{figure*}
\centerline{
\psfig{figure=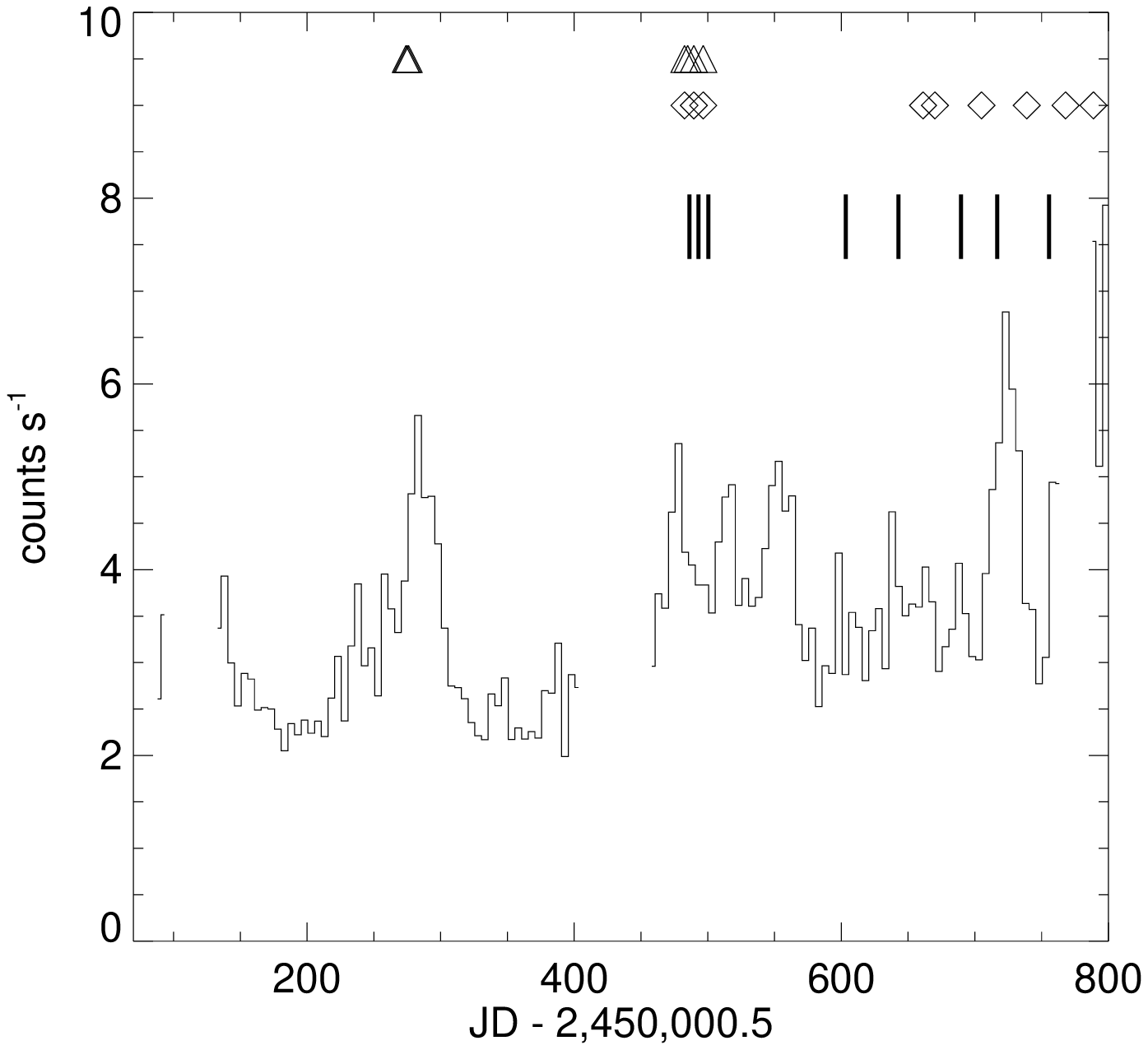,width=0.45\textwidth} 
\psfig{figure=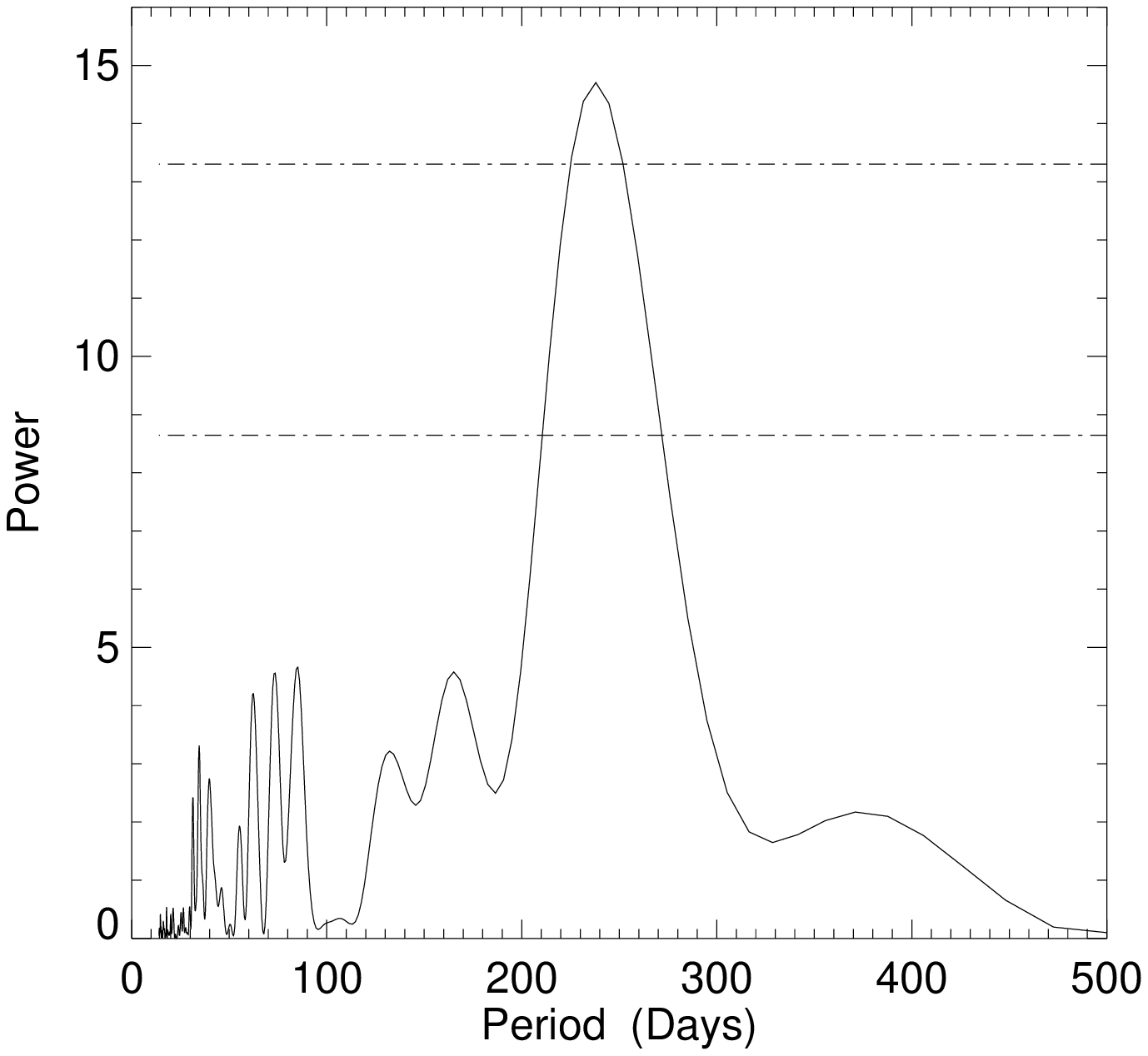,width=0.45\textwidth} }
\caption{\small \emph {Left:} RXTE All Sky Monitor data for GX~339--4 (5 day
  averages in the 1.3--12.2\,keV band) vs. Truncated Julian Date (TJD)
  $\equiv$ Julian Date (JD) $-2450000.5$.  Dashes indicate dates of our
  RXTE pointed observations, diamonds indicate dates of MOST observations
  (Hannikainnen et al. 1998), and triangles indicate dates of ATCA radio
  observations (Corbel et al.  1998).  \emph{Right:} A Lomb-Scargle
  Periodogram (cf., Lomb 1976, Scargle 1982) of the ASM data for TJD
  $<800$. We have used 600 periods ranging from 2 weeks to 500 days.  Lines
  are estimates of the 99.9\% and 90\% significance levels. \label{fig:asm}}
\end{figure*}

We used data from the All Sky Monitor on RXTE to study the long-term
behavior of GX~339-4. The ASM is an array of three shadow cameras combined
with position sensitive proportional counters that provides for a
quasi-continuous coverage of the sky visible from RXTE.  In practice,
lightcurves in three energy bands --- 1.3--3.0\,keV, 3.0--5.0\,keV, and
5.0--12.2\,keV --- as well as over the whole ASM band are publically
available from the ASM data archives (\cite{lochner:97a}).  Typically there
are several 90\,s measurements available for each day.  Further
descriptions of the instrument and first year results are presented by
\citey{levine:96a} and \citey{remillard:97a}.

In Figure~\ref{fig:asm} we present the ASM data of GX~339--4 up until
Truncated Julian Date (TJD) $\approx 800$ (1997 November 15).  See paper I
for a presentation of the complete to date ASM data of GX~339--4. We also
indicate in this figure the dates of our RXTE observations, as well as the
dates of Australian Telescope Compact Array (ATCA) and Molongolo Observing
Synthesis Telescope (MOST) radio observations of GX339--4
(\cite{fender:97a,corbel:98a,hanni:98a}; paper I).  The average $1\sigma$
error bar for a point presented in Figure~\ref{fig:asm} is $\approx
0.4~{\rm cts~s^{-1}}$. Note the three peaks (at TJD $\approx 280$,
480--580, 780) in the ASM data that occur at intervals separated by
$\approx 250$\,days.

We determined the significance of any possible long term periodicities in
the ASM light curves by computing the Lomb-Scargle Periodogram
(\cite{lomb:76a,scargle:82a}) for the 1.3--12.2\,keV band for 5-day average
data. Here we only average data where the best fit to the source position
and flux in an ASM observation has a $\chi_{\rm red}^2 \le 1.5$ (cf.
\cite{lochner:97a}) in \emph{each} of the three ASM energy channels.  We
only considered ASM data taken before TJD 800, as GX~339--4 underwent a
state change shortly thereafter (see paper I).  The periodogram shown in
Figure~\ref{fig:asm} reveals evidence of a 240 day period at greater than
the 99.9\% significance level. [The significance levels were estimated
following the methods of \citey{horne:86a}.]  Epoch folding (cf.
\cite{leahy:83a,schwarz:89a,davies:90a}) of the ASM lightcurves also shows
evidence of this 240 day periodicity.

Long timescale periodicities and quasi-periodicities are relatively common
in ASM observations of binary sources (Remillard 1997, Private
Communication).  Evidence for a 294\,d periodicity in Cygnus~X--1 has been
previously reported (\cite{kemp:83a,priedhorsky:83a}), and is readily
apparent in the ASM data during the hard state.  A 198\,day periodicity
also has been observed in LMC~X--3 (\cite{cowley:91a,wilms:98a}).  We note
that the observed timescales of these periodicities are comparable to the
timescales expected from the the radiation pressure driven warping
instability discovered by \citey{pringle:96a} (see also
\cite{maloney:96a,maloney:97a,maloney:98a}).  This is a fairly generic
instability that causes a radiatively efficient (i.e., non-advection
dominated) accretion disk to warp and precess on ${\cal O}(100~{\rm day})$
timescales. Such a timescale is consistent with the periodicities seen in
many binaries and with the 240 periodicity that we see in GX~339--4.

As discussed in paper I, however, both the Compton corona models and the
ADAF models of the spectral data suggest that the observed flux variations
are in large part attributable to variations of the coronal radius.  A pure
warped, precessing disk model would invoke only inclination angle effects.
This is unlikely to be the case for GX~339--4.  As discussed by
\citey{maloney:96a}, however, the precession timescale and warp shape are
more sensitive to the outer boundary conditions than the inner boundary
conditions.  Perhaps then it is possible that what we observe is a
combination of a quasi-steadily precessing disk on large radii combined
with coronal structure changes on small radii.  Although 240 days is a
characteristic timescale in both the Lomb-Scargle periodogram and in the
epoch folding analysis, it is obvious from Figure~\ref{fig:asm} that we are
not observing a strictly periodic phenomenon.

\section{Power Spectral Densities}\label{sec:psd}

We studied the variability of GX~339--4 using data from the Proportional
Counter Array (PCA) on-board RXTE.  The PCA consists of five nearly
identical co-aligned Xenon proportional counter units (PCUs) with a total
effective area of about 6500\,cm$^{2}$, and it is sensitive in the energy
range from 2\,keV to $\sim 60$\,keV (\cite{jahoda:96b}).  We only used data
where all five PCUs were turned on, and where the elevation angle between
the spacecraft pointing direction and the limb of the earth was greater
than $10^\circ$.  PCA count rates for the GX~339--4 observations ranged
from 200 to 800~${\rm cts~s^{-1}}$, and were of ${\cal O}$(10\,ks)
duration.

To study the short timescale variability, we created $2^{-8}$\,s resolution
lightcurves in 4 different energy bands: 0--3.9\,keV, 3.9--7.5\,keV,
7.5--10.8\,keV, 10.8--21.9\,keV; hereafter labelled bands A, B, C, and
D\footnote{We also created $2^{-13}$\,s resolution lightcurves to search
  for signatures of high frequency features.  No evidence for such features
  was found. Note also that the effective lower limit for energy band A is
  more $\approx 2$\,keV.}.  Note that this is one fewer energy band and
half the time resolution than for our Cyg X--1 observations
(\cite{nowak:98a}).  This was required to obtain good statistics because
the GX~339--4 observations discussed here ranged from 200--800\,${\rm
  cts~s^{-1}}$, as opposed to the 4500\,${\rm cts~s^{-1}}$ observed for
Cyg~X--1.  Energy bands A--D each had roughly the same count rate for a
given observation.  As for the spectral analysis (paper I), we found that
all the observations were similar in their properties, with the exception
of Observation 5. Observation 5 was approximately a factor of 5 fainter
than the brightest observation (Observation 1).

\begin{figure*}
\centerline{ \psfig{figure=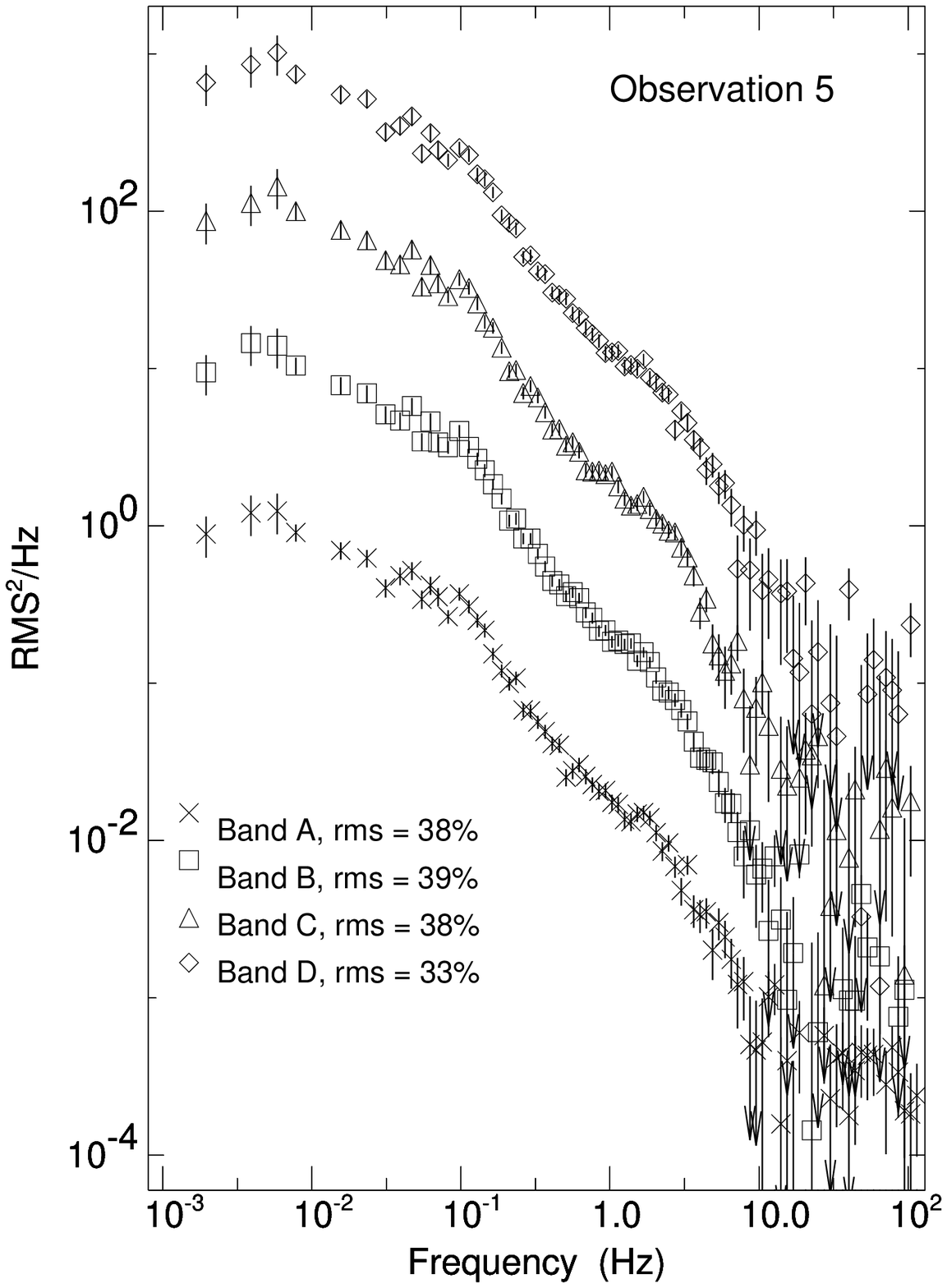,width=0.45\textwidth} ~
\psfig{figure=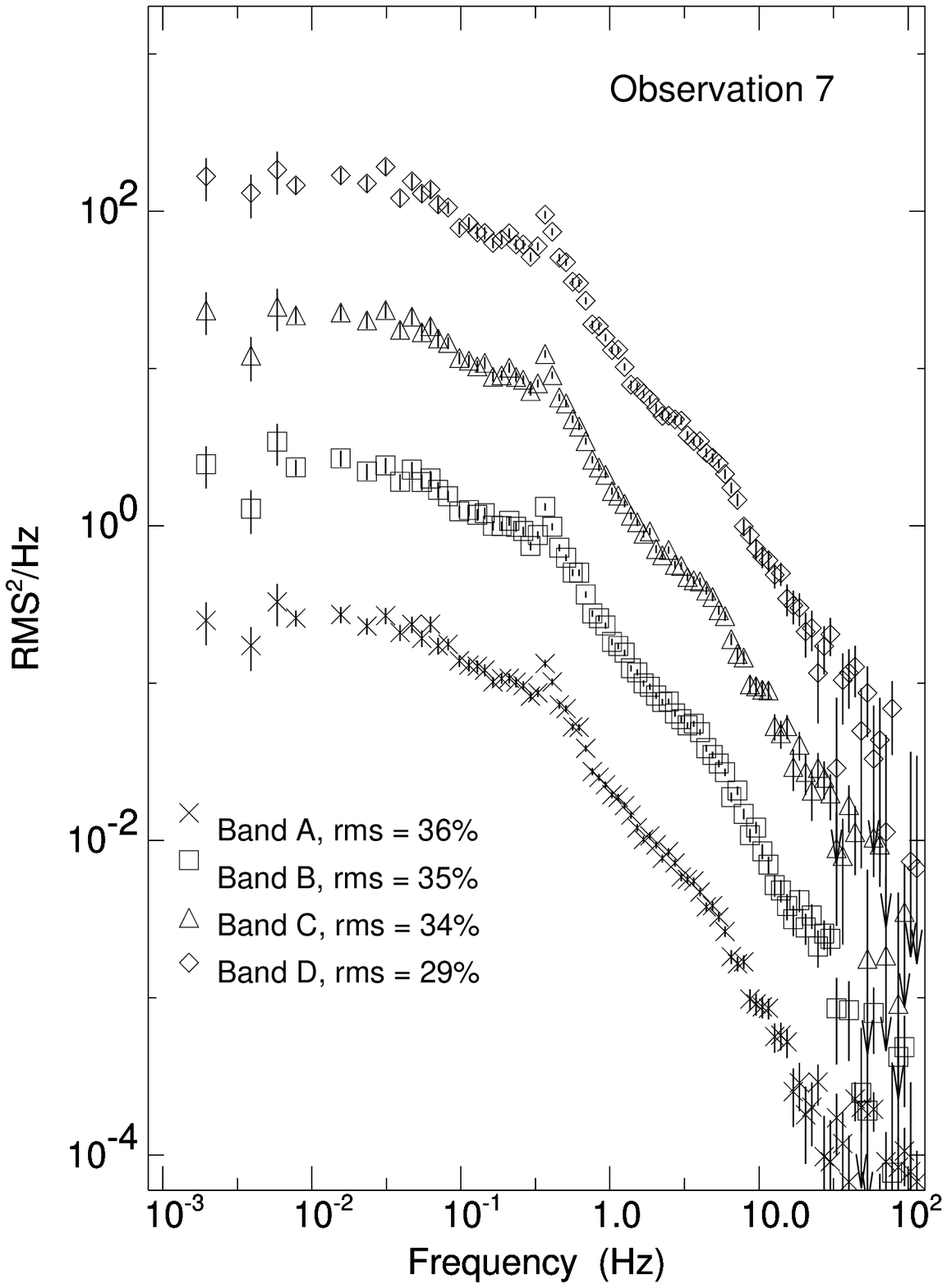,width=0.45\textwidth} }
\bigskip
\caption{\small PSDs with associated uncertainties for Observation 5 ({\it
    left}) and for Observation 7 ({\it right}).  All PSD are for the
  one-sided normalization of Belloni \& Hasinger (1990b), where integrating
  over positive frequencies yields the total mean square variability.
  Crosses correspond to energy band A, squares to energy band B shifted
  upwards by a factor of 10, triangles to energy band C shifted upward by a
  factor of 100, and diamonds to energy band D shifted upward by a factor
  of 1000.  (See text for the energy range of the bands.) Root mean square
  (rms) variability values were calculated between $f= 7 \times
  10^{-3}$--$40$\,Hz. \label{fig:psds}}
\end{figure*}

As for our RXTE observations of Cyg~X--1 (\cite{nowak:98a}), we combined
low frequency ($\approx 8\times10^{-4}$--$7\times10^{-3}$\,Hz) FFTs made
from a few (6--16) data segments of length 512--1024\,s with higher
frequency ($7\times10^{-3}$--128\,Hz) FFTs made from somewhat more
($\aproxgt 70$) data segments of length 128\,s.  From these data segments
we created PSDs for all of our observations. We present two examples of the
resulting PSDs in Figure~\ref{fig:psds}.  With the exception of Observation
5, all PSDs were qualitatively and quantitatively similar to that presented
in Figure~\ref{fig:psds} for Observation 7.  The PSDs for Observations 1--4
and 6--7 have shapes roughly similar to what we observed for Cyg~X--1: flat
from $\approx 10^{-3}$--$0.03$\,Hz, approximately $\propto f^{-1}$
inbetween 0.03--3\,Hz, and steeper above $\approx 3$\,Hz
(\cite{nowak:98a}).  The PSDs for Observation 5 have qualitatively similar
shapes; however, the break frequencies are approximately a factor of three
lower.

\begin{figure*}
\centerline{
\psfig{figure=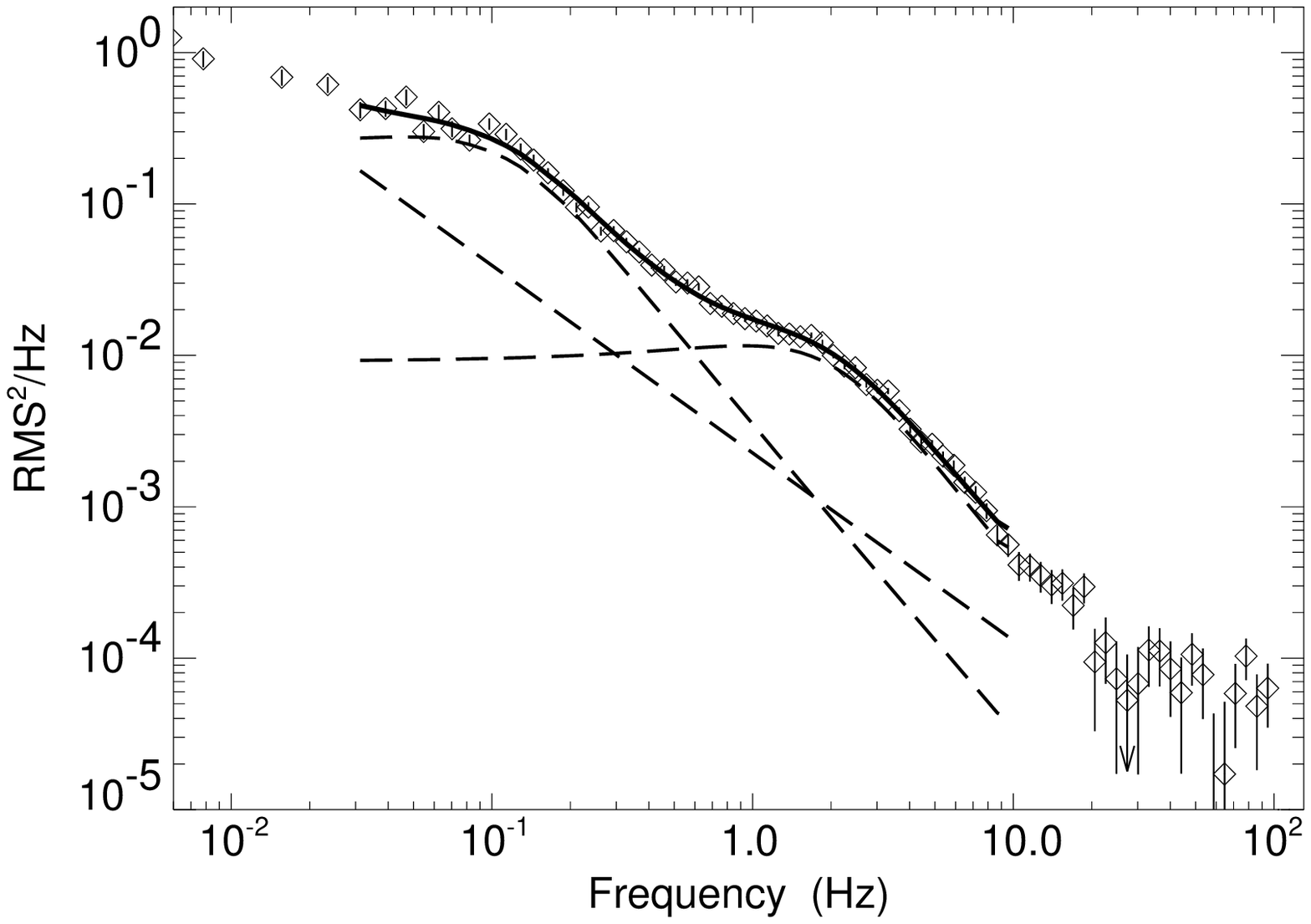,width=0.45\textwidth} ~
\psfig{figure=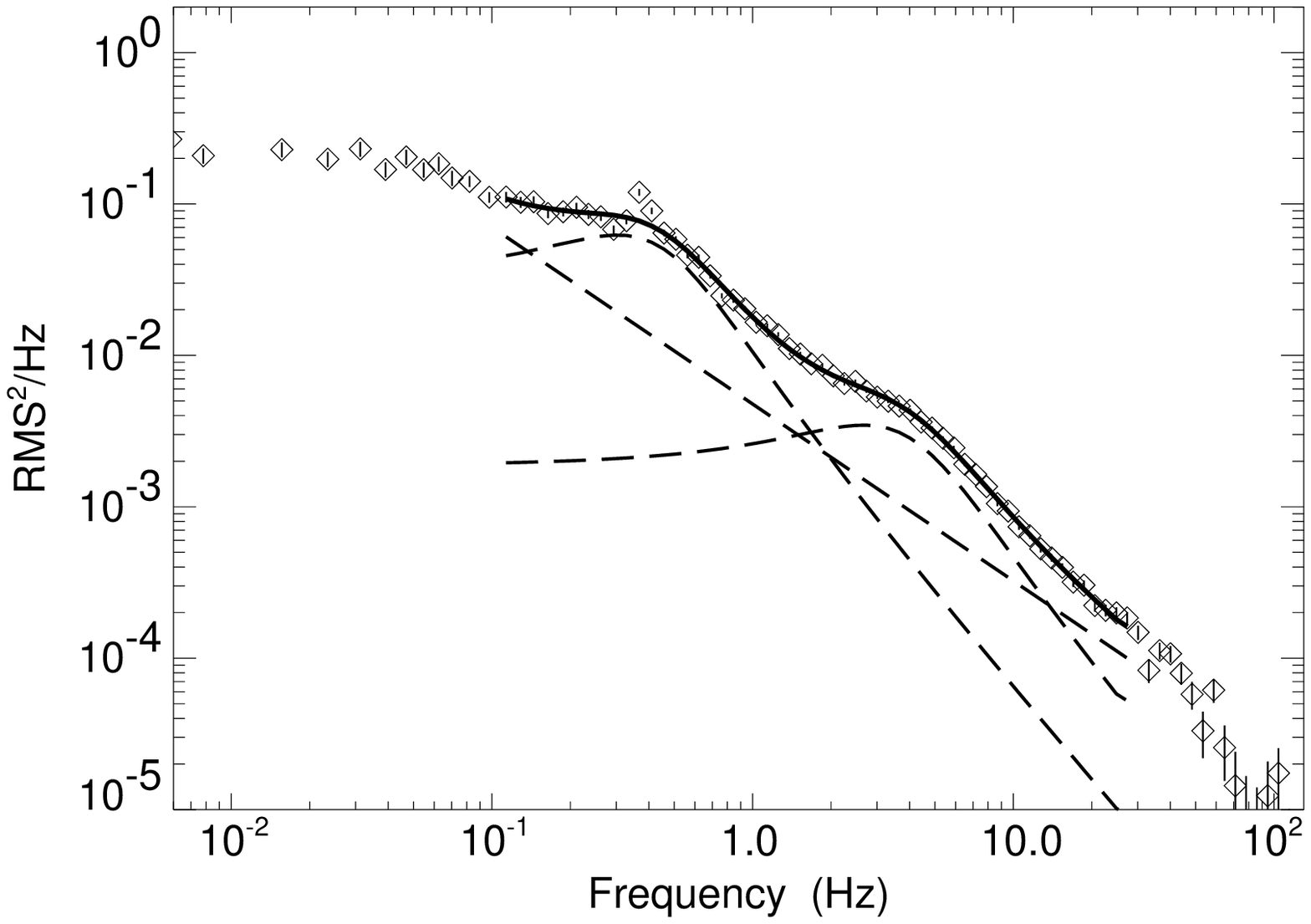,width=0.45\textwidth} 
}
\bigskip
\caption{\small {\it Left:} Diamonds are the PSD for Observation 5 (energy
  bands A--D summed), normalized as in Belloni \& Hasinger (1990b). {\it
    Right:} Diamonds are the PSD for Observation 7 (energy bands A--D
  summed).  For both figures, the solid line is the best fit power law plus
  two broad Lorentzians.  (Fit range is 0.03--10\,Hz for Observation 5, and
  0.1--30\,Hz for Observaion 7.)  Dashed lines show the individual
  components of the fits.}
\label{fig:psdfits}
\end{figure*}

Doubly broken power laws provided adequate descriptions of the PSDs for our
RXTE observations of Cyg~X--1 (\cite{nowak:98a}).  Here, however, we
clearly see that the GX~339--4 PSDs contain more structure.  An $\approx
0.3$\,Hz quasi-periodic oscillation (QPO) is evident in Observation 7.  In
fact, this QPO appears at some level in \emph{all} of our observations
\emph{except} Observation 5.  Even ignoring the QPO features, the PSD is
only marginally approximated by a doubly broken power law.

We have tried to fit the shape of the PSDs with a number of different
functional forms.  For example, a singly broken power law with a Lorentzian
\emph{absorption} feature at $\approx 0.3$--$1$\,Hz provides good fits to
all of the PSDs.  Here we show the results of fitting a weak power law
plus two broad Lorentzians.  As we further elaborate upon in \S\ref{sec:lag}
and \S\ref{sec:discuss}, such a fit may have some physical relevance.  The
Lorentzian and power law components may represent separate, broad-band
processes that are individually coherent (between their soft and hard
variability), but that are incoherent with one another.

Sample fits to the broad-band power are presented in
Figure~\ref{fig:psdfits}, and results for all of our data are presented in
Table~\ref{tab:tqpo}.  The functional form of the Lorentzians that we fit
is given by
\begin{equation}
P(f) = \pi^{-1} \frac{R^2 Q f_0}{f_0^2 + Q^2 (f - f_0)^2} ~~.
\label{eq:lorentz}
\end{equation}
Here $f_0$ is the resonant frequency of the Lorentzian, $Q$ is the quality
factor ($\approx f_0/\Delta f$, where $\Delta f$ is the
full-width-half-maximum of the Lorentzian), and $R$ is the fit amplitude
(root mean square variability, rms $= R [ 1/2 - \tan^{-1}(-Q)/\pi]^{1/2}$,
i.e. rms $=R$ as $Q \rightarrow \infty$).  For all the fits presented in
Table~\ref{tab:tqpo}, $Q \aproxlt 1$. We thus cannot consider the
broad-peaks in the observed PSDs to be ``quasi-periodic oscillations''.
The fits are more rightly considered to be indicative of broad-band power
with characteristic timescales $\sim f_0^{-1}$.  Note also that, for all of
the fits, $\chi^2_{\rm red} \aproxlt 2$--3.  Although this formally
represents an unacceptable fit, this is partly attributable to the
excellent statistics achievable with RXTE.  Even small fractional
deviations from the fit are highly statistically significant.  In practice
we have found that $\chi^2_{\rm red} \aproxlt 2$ is nearly impossible to
achieve with any simple functional fits.  Futhermore, the fits with the
largest $\chi^2_{\rm red}$ showed the most pronounced 0.3\,Hz QPO, as in
Observation 7.

\begin{table*}
\caption{\small Fits to the PSD of the form: ${\rm A} f^\Gamma + \pi^{-1} 
\left( {{\rm R}_1^2 Q_1 f_1}/{[ f_1^2 + Q_1^2 (f - f_1)^2]} + 
{{\rm R}_2^2 Q_2 f_2}/{[ f_2^2 + Q_2^2 (f - f_2)^2]} \right )$.  
Energy ranges of bands A--D are described
in the text.  T represents energy bands A--D summed together.  All fits
were in the range $f=0.1$--30\,Hz (47 degrees of freedom), except for
Observation 5 which was fit in the
range $f=0.03$--10\,Hz (45 degrees of freedom). Errors are the nominal 
90\% confidence level for one interesting parameter 
($\Delta \chi^2 = 2.71$).}  \label{tab:tqpo}
\bigskip
{\small
\input{tqpo_pow} }
\end{table*}

\begin{figure*}
\label{fig:psdstuff}
\centerline{
\psfig{figure=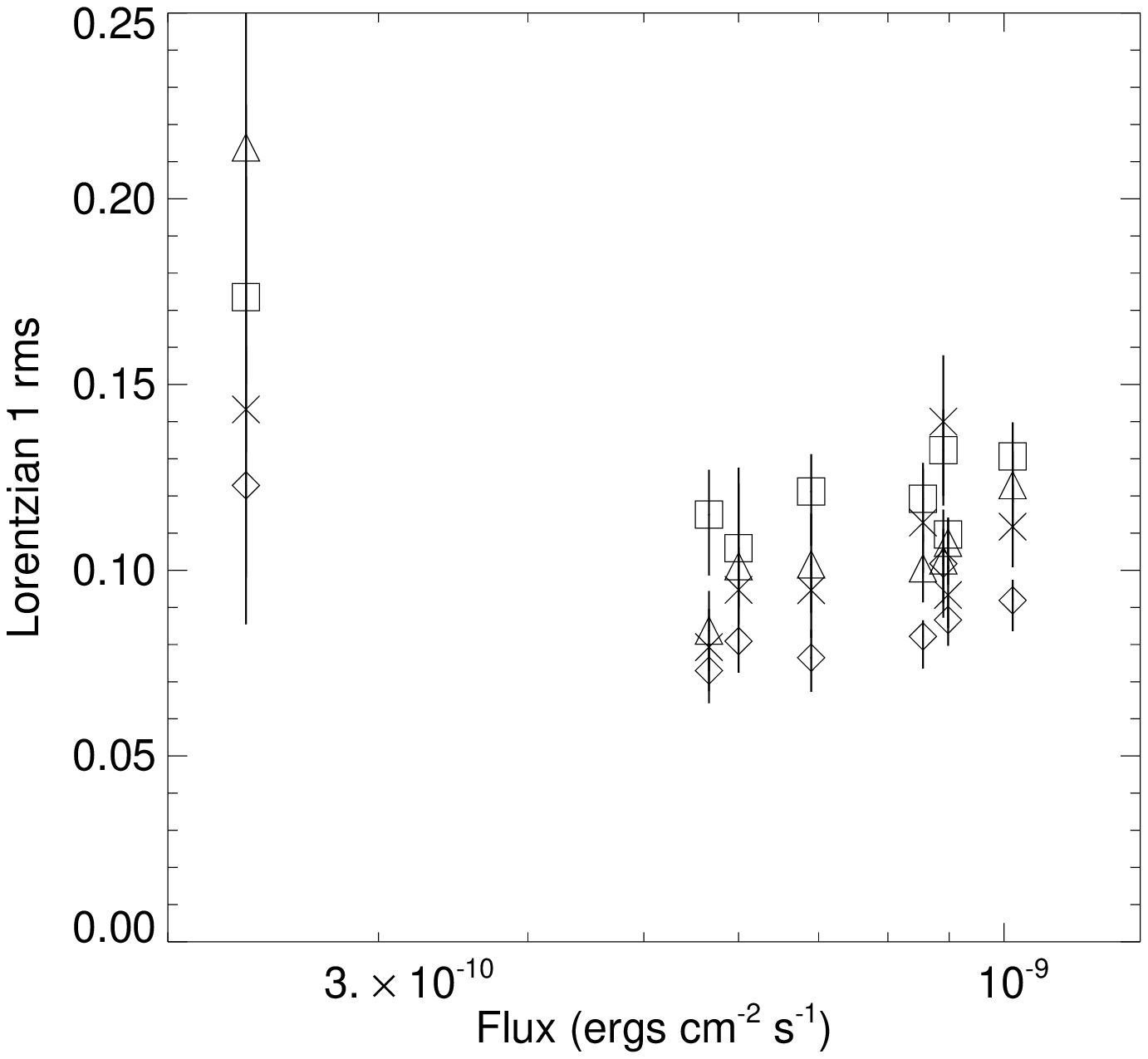,width=0.45\textwidth} ~
\psfig{figure=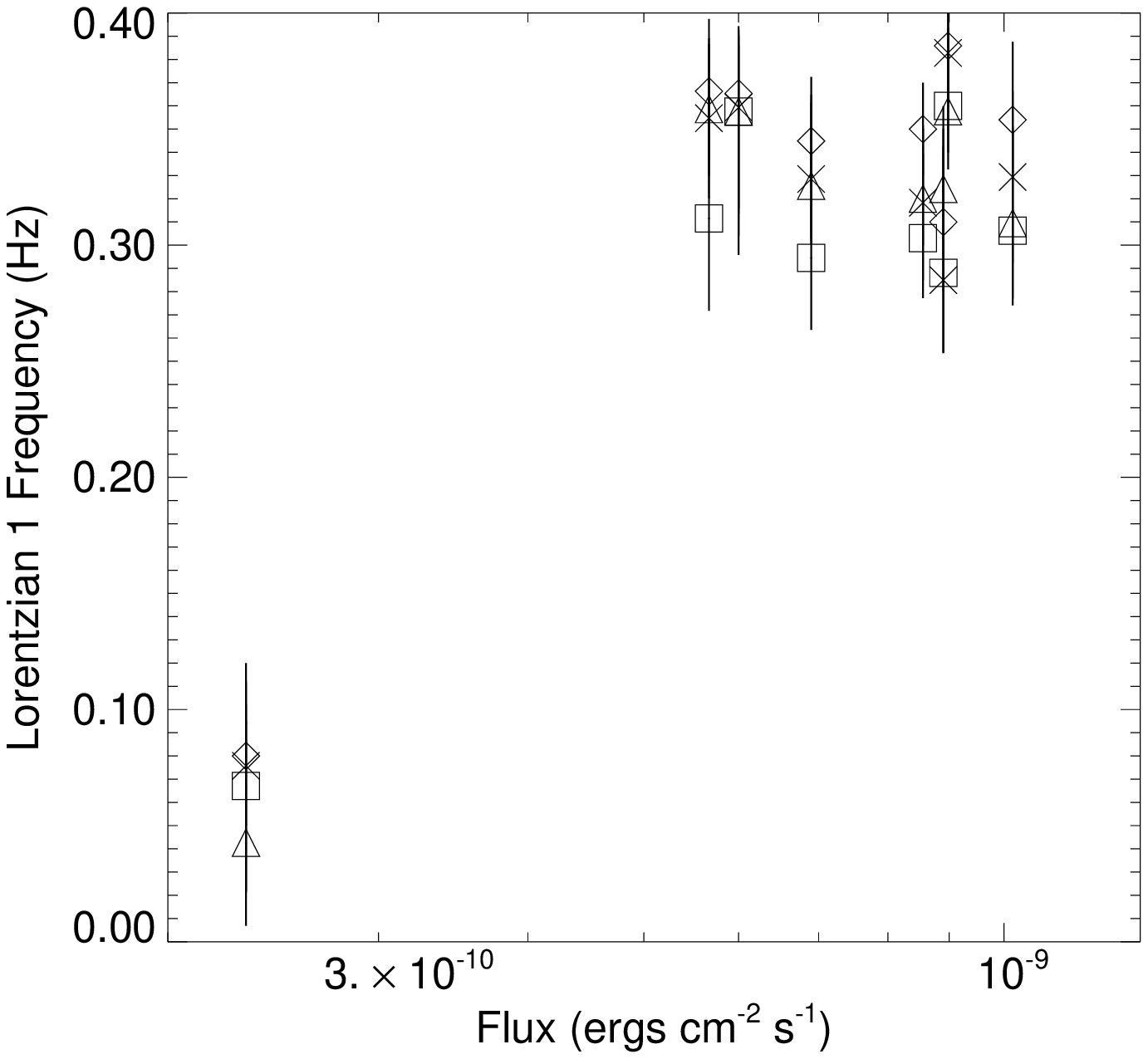,width=0.45\textwidth} }

\centerline{
\psfig{figure=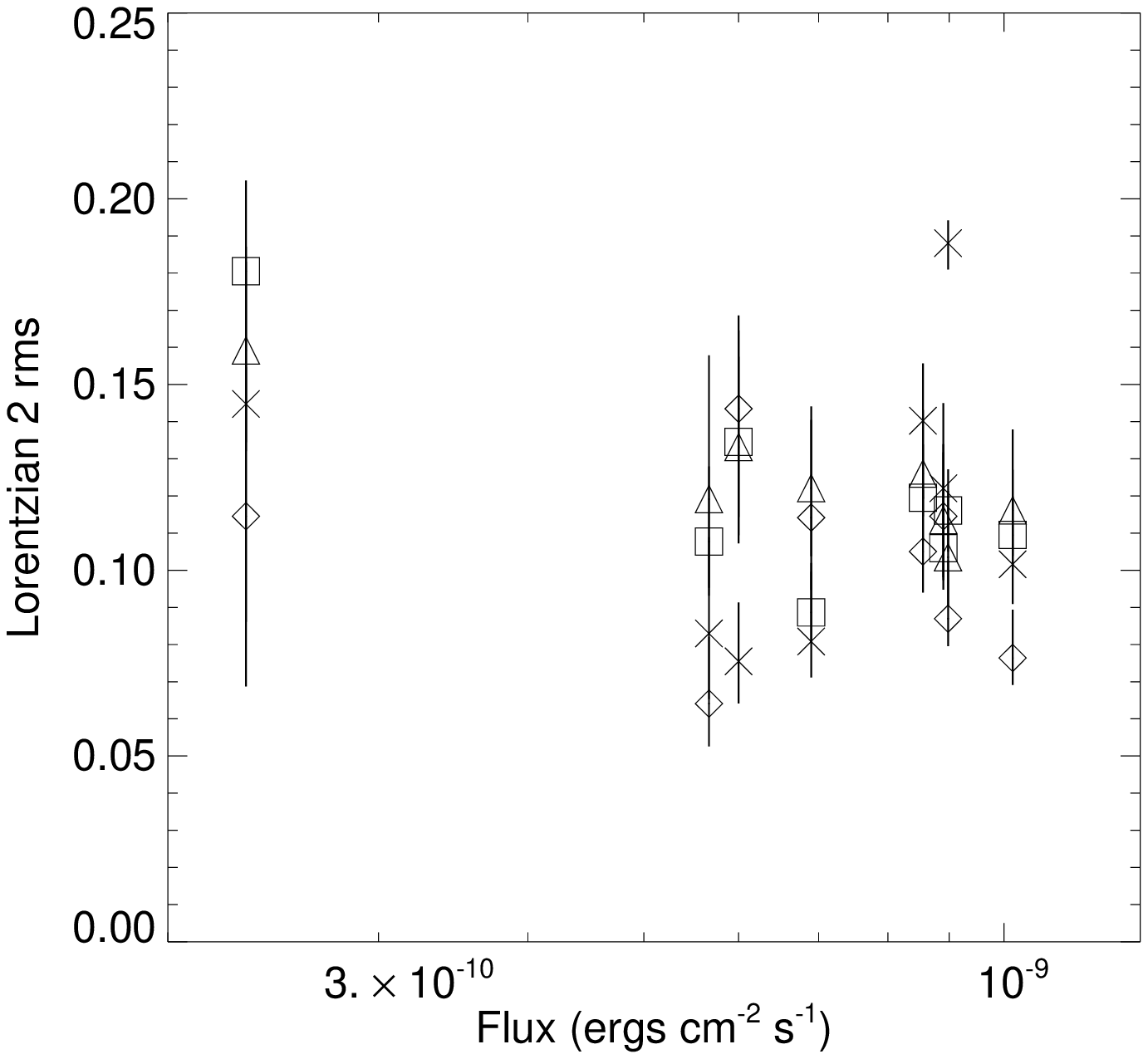,width=0.45\textwidth} ~
\psfig{figure=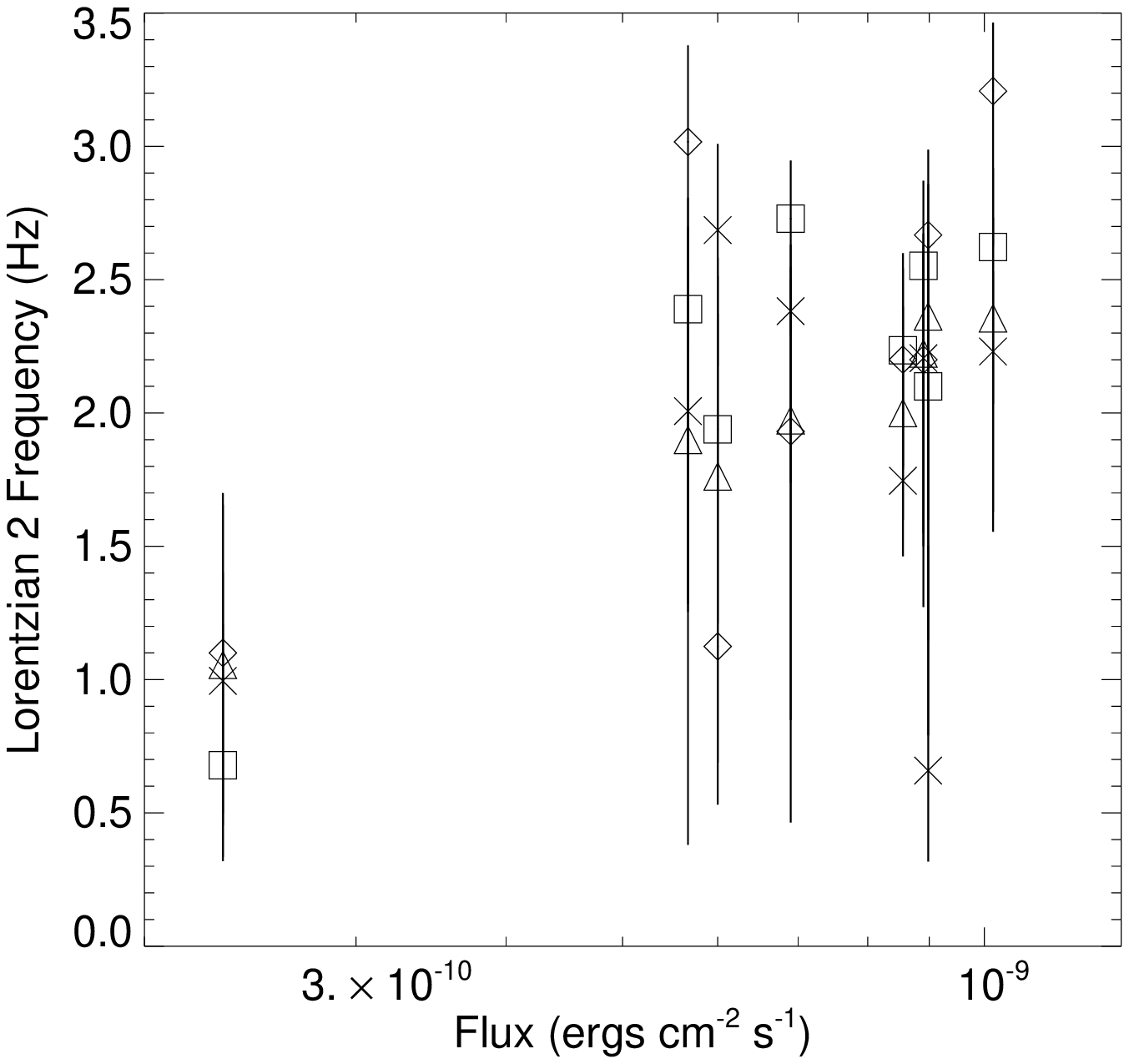,width=0.45\textwidth} 
}
\bigskip
\caption{\small {\it Upper Left:} rms variability of the low-frequency broad
Lorentzian fit component vs. the 3--9\,keV flux of the observation.
{\it Upper Right:} Peak frequency of the low-frequency broad
Lorentzian fit component vs. the 3--9\,keV flux of the observation.
{\it Lower Left:} rms variability of the high-frequency broad
Lorentzian fit component vs. the 3--9\,keV flux of the observation.
{\it Lower Right:} Peak frequency of the high-frequency broad
Lorentzian fit component vs. the 3--9\,keV flux of the observation.
Symbols are the same as for Figure 2.}
\end{figure*}

For Observations 1--4 and 6--7, the low frequency broad-band power was
peaked near $\approx 0.3$\,Hz, while the high frequency broad-band power
was peaked near $\approx 2.5$\,Hz.  The rms variability amplitudes were all
comparable for these observations.  Observation 5 had a somewhat larger rms
variability amplitude.  The trend for all observations was for the rms
variability amplitude, measured from $7 \times 10^{-3}$--40\,Hz, to
decrease with increasing energy band, although this is only marginally
evident in the individual fit components (cf. Figure~\ref{fig:psdstuff}).
In Figure~\ref{fig:psdstuff} we plot the best fit rms variabilities and
peak frequencies for the two Lorentzian fit components vs. the observed
3--9\,keV flux.

Observations 1--4 and 6--7 span a factor of two in observed 3--9\,keV flux.
The PSD parameters, however, show no obvious trends with flux, except for
exhibiting weak evidence for the rms variability of the low frequency
Lorentzian fit component to increase with 3--9\,keV luminosity.  Perhaps
the most remarkable aspect of this subset of the observations is how
similar all the PSDs appear to one another despite the factor of two spread
in the observed 3--9\,keV flux.

\begin{figure*}
\centerline{
\psfig{figure=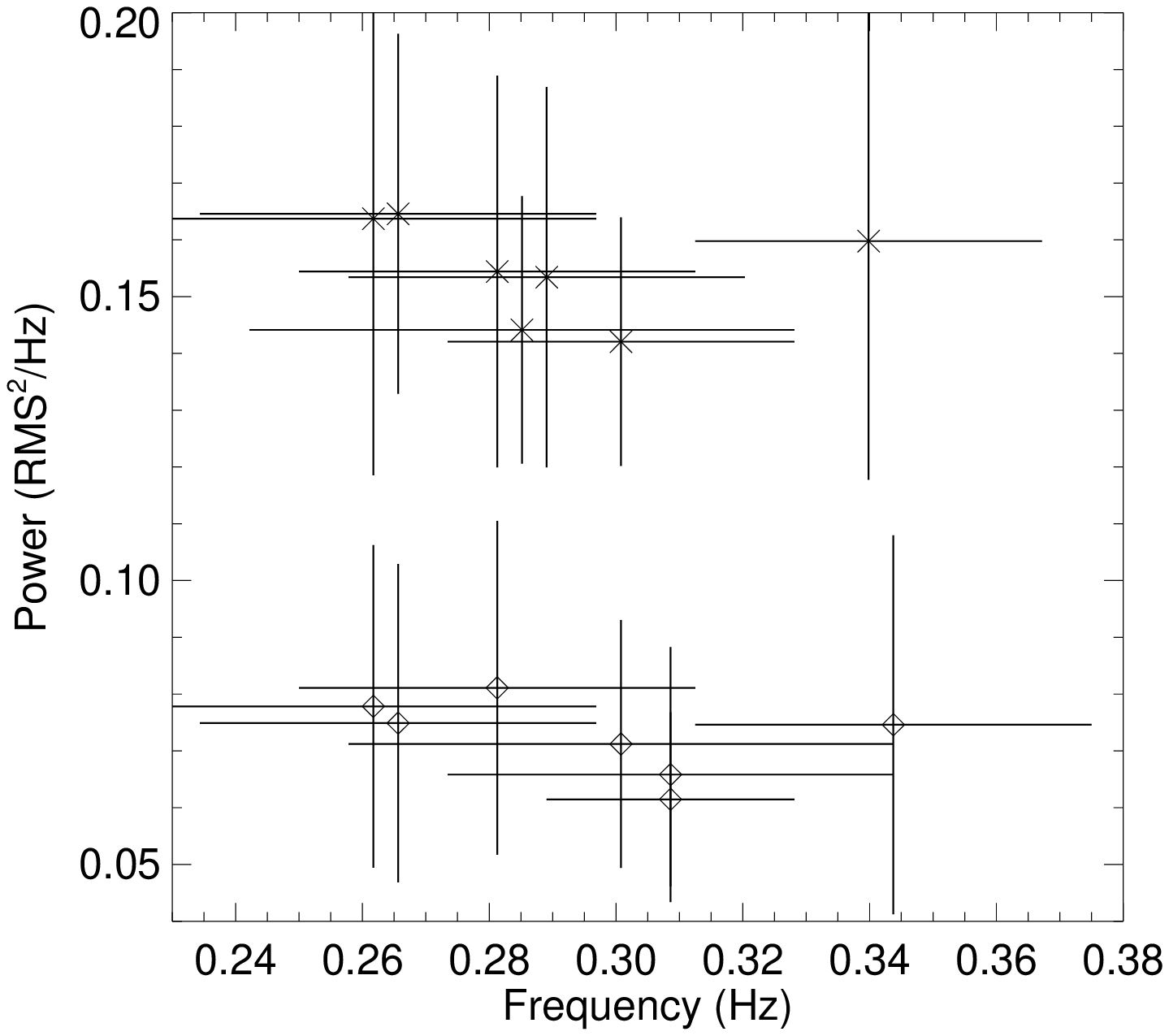,width=0.32\textwidth} 
}
\smallskip
\caption{\small 
  QPO PSD amplitude vs. frequency.  The bars extend from the values for the
  dip to the values of the peak (see text), and the points are placed at
  the midpoint.  Crosses are for energy band A and diamonds are for energy
  band D, with PSD amplitude lowered by 0.1. \label{fig:qpostuff}}
\end{figure*}

The $\approx 0.3$\,Hz QPO is most prominent in Observation 7.  Its strength
and width are somewhat difficult to characterize, however, as it is
difficult to determine the `continuum' level of the PSD to which it should
be compared.  The rise of the QPO appears sharper than its decline, and
typically there is a PSD `dip' at $\approx 0.05$\,Hz before the peak of the
QPO.  We have tried a variety of narrow Lorentzian plus broad-band power
fits to the features; however, the best fit parameters are highly dependent
upon the assumed form of the underlying broad-band PSD.  Furthermore,
narrow Lorentzian fits do not capture the asymmetric shape of the QPO.

Being unable to find a satisfactory functional fit to the QPO, we have
chosen to characterize it in the following manner.  We have measured the
location of the PSD `dip' that occurs before the QPO peak, as well as the
location of the QPO peak. (Factoring in noise fluctuations, the QPO is
sharp enough from dip to peak that the location of each is likely accurate
to better than two unaveraged frequency bins, i.e. 0.015\,Hz.)  We have
also measured the dip and peak PSD amplitudes.  As for the fits to the
broad-band PSD discussed above, there are no obvious trends between QPO
parameters and the observed 3--9\,keV flux.  Furthermore, there is no
obvious trend between QPO peak power and frequency, as shown in
Figure~\ref{fig:qpostuff}.  Although the QPO frequency is
characteristically near 0.3\,Hz, it is not at a steady frequency and it
varies over a range of 0.08\,Hz. Peak QPO amplitudes and widths, however,
are somewhat less variable.  Note that the \emph{total} rms variability in
the region of the QPO typically is $\approx 5\%$, therefore the QPO
amplitude is at most a few percent.

\section{Coherence and Time Lags}\label{sec:lag}

\begin{figure*}
\centerline{
\psfig{figure=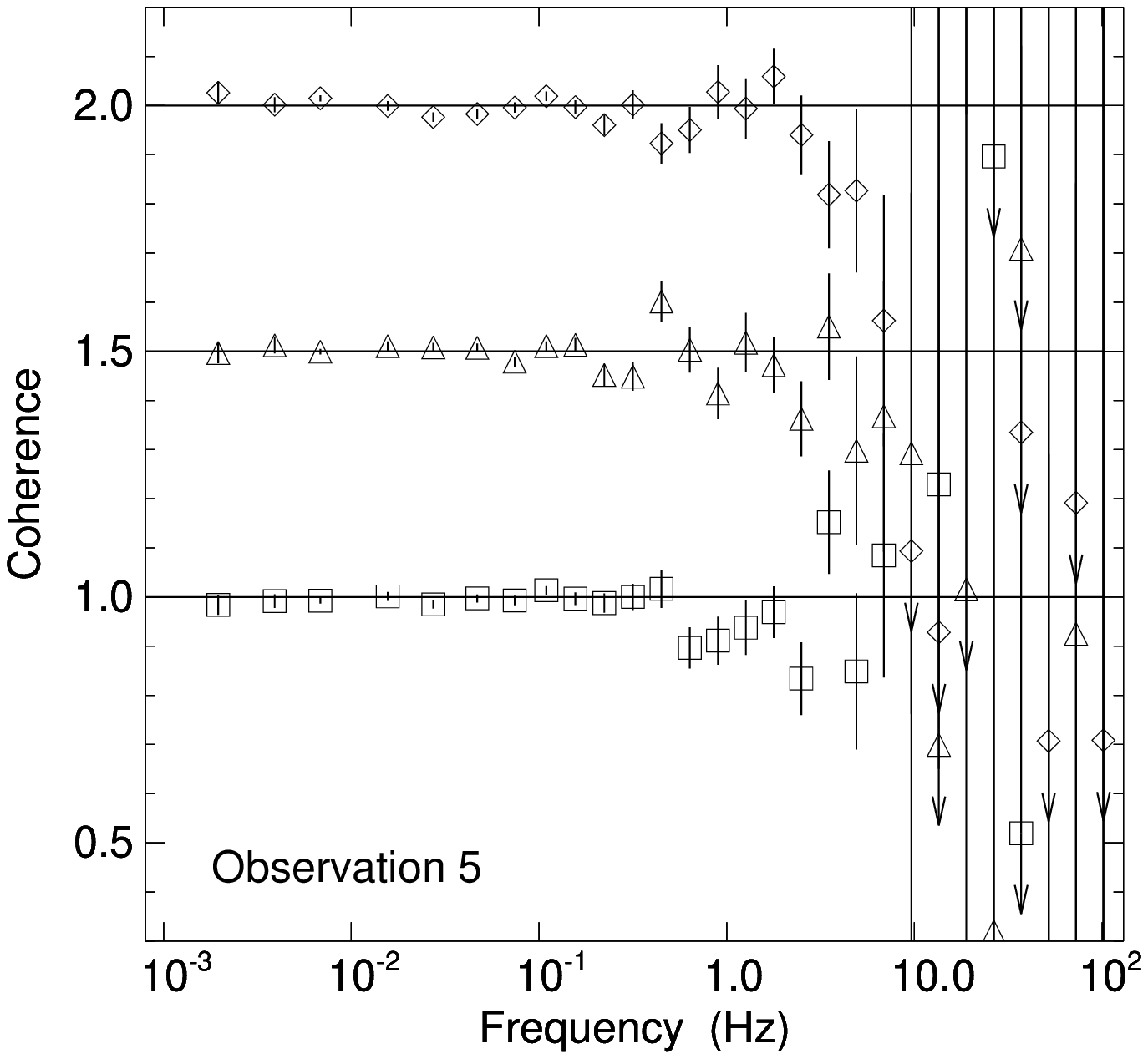,width=0.45\textwidth} ~
\psfig{figure=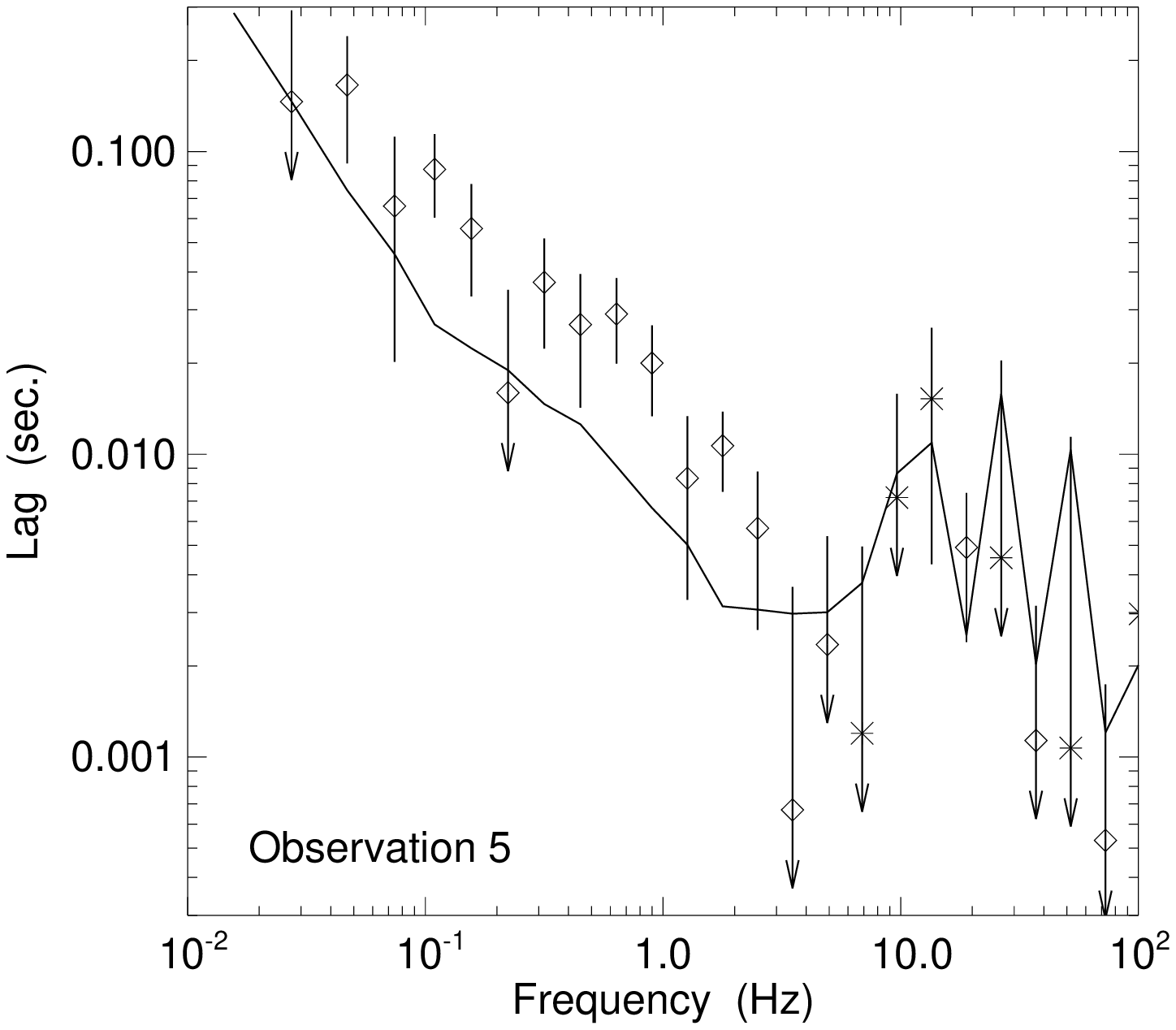,width=0.45\textwidth} 
}
\centerline{
\psfig{figure=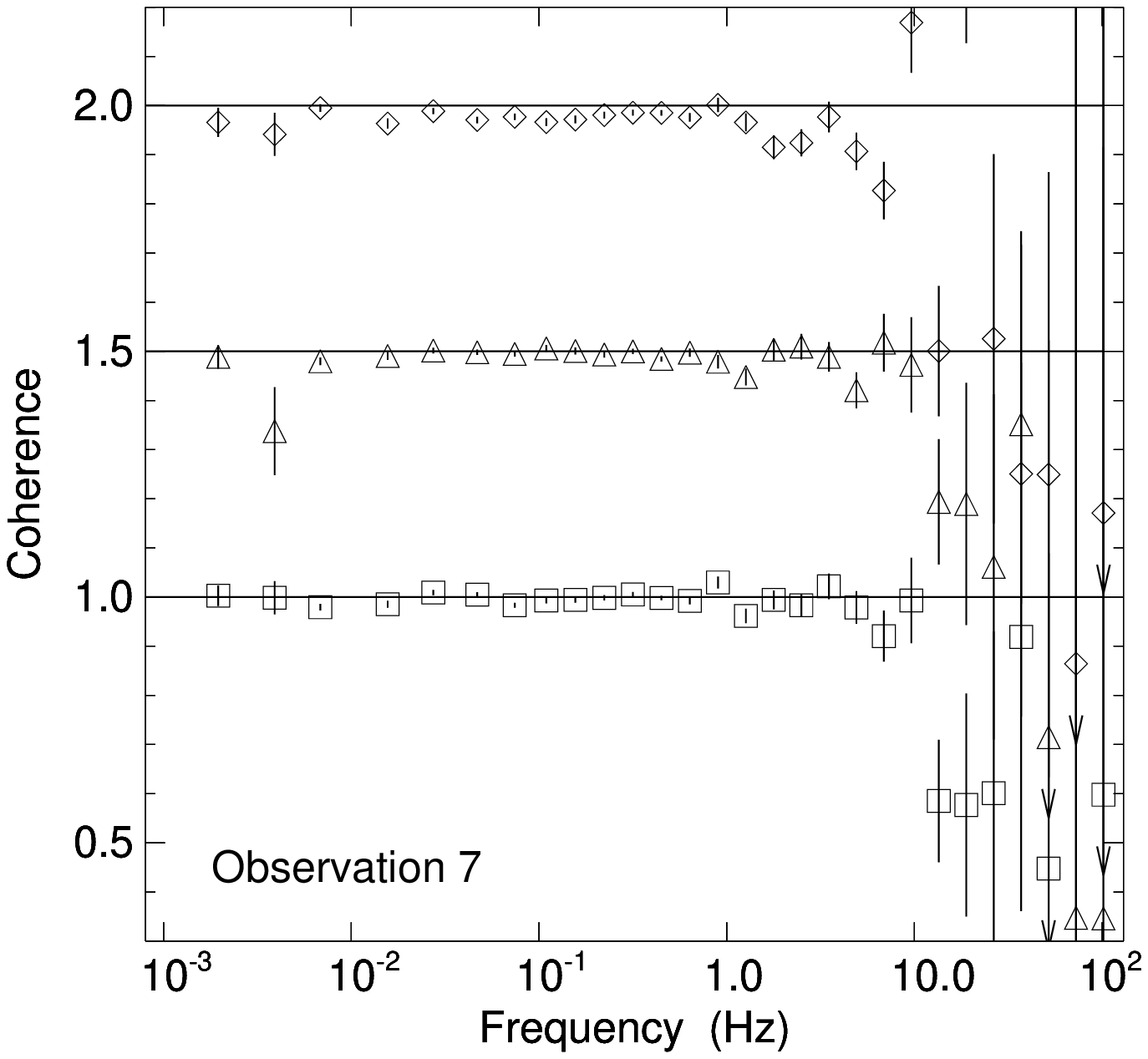,width=0.45\textwidth} ~
\psfig{figure=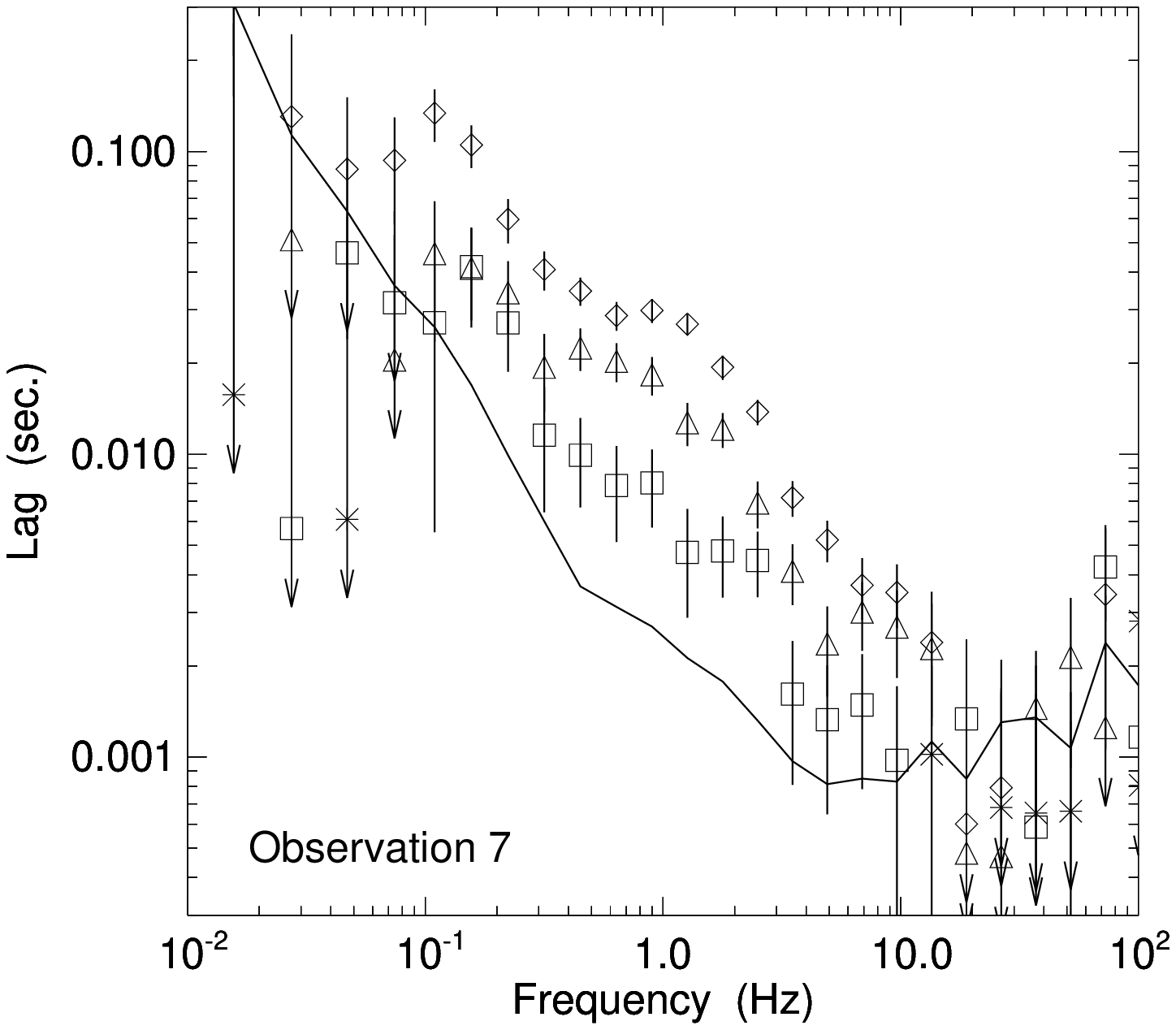,width=0.45\textwidth} 
}
\bigskip
\caption{\small Coherence function and time lags of various energy channels
[squares: channel B, triangles: channel C (coherence offset by 0.5),
diamonds: channel D (coherence offset by 1.0)] as compared to energy
channel A. {\it Top Left:} Coherence for Observation 5.  Solid lines
correspond to unity coherence.  {\it Top Right:} Time lags for Observation
5.  Crosses are where the soft variability lags the hard variability.
Solid line is the Poisson noise level for channel D time lags.  {\it Bottom
Left:} Coherence for Observation 7.  Solid lines correspond to unity
coherence.  {\it Bottom Right:} Time lags for Observation 7.  Crosses are
where the soft variability lags the hard variability.  Solid line is the
Poisson noise level for channel D time lags.}
\label{fig:cross}
\end{figure*}

As discussed by \citey{vaughan:97a}, the coherence function is a measure of
the degree of linear correlation between two time series. Specifically, it
gives the fraction of the mean-squared variability at a given Fourier
frequency in one time series that can be attributed to, or equivalently
predicted from, the other.  The fact that coherence is often near unity
over a wide range of frequencies (cf. \cite{vaughan:97a,nowak:98a}) is what
makes it meaningful to then talk about characteristic Fourier
frequency-dependent time delays between two time series (cf. 
\cite{miyamoto:89a,miyamoto:92a,vanderklis:89b,nowak:98a}).

The techniques that we used to calculate the coherence and time delays for
the GX~339--4 lightcurves are discussed in \citey{vaughan:97a} and
\citey{nowak:98a}.  For all observations we have calculated the coherence
function,
\begin{equation}
\gamma^2(f) = \frac{ | \langle S^*(f) H(f) \rangle |^2 }{
     \langle | S(f) |^2 \rangle \langle | H (f) |^2 \rangle } ~~,
\label{eq:cohdef}
\end{equation}
between the FFTs for energy band A [$S(f)$] and the FFTs for the other
three energy bands [$H(f)$, angle brackets indicating averages over Fourier
frequencies and individual data segments; cf.
\cite{vaughan:97a,nowak:98a}].  For all observations except for Observation
5, we have calculated the time-delay between energy band A and the other
three energy-bands.  For Observation 5, noise limitations only allow us to
calculate the time-delay between the lightcurves for energy band D and
energy band A.  We have averaged over logarithmically spaced frequency
bins, ranging over frequencies $f\rightarrow 1.4 f$ for all observations.

As shown in Figure~\ref{fig:cross}, the coherence function is near unity
from $10^{-3}$ to $\approx 3$\,Hz (Observation 5) or $\approx 10$\,Hz (all
other observations). Above $\approx 3$--10\,Hz, there is a noticeable drop
in coherence, similar to what we saw with our RXTE observations of Cyg~X--1
(\cite{nowak:98a}). The Cyg~X--1 observations also showed a loss of
coherence below $\approx 0.02$\,Hz; however, the coherence in GX~339--4
remains near unity down to Fourier frequencies as low as $\approx
10^{-3}$\,Hz.  

We note that the coherence between energy bands A and D shows evidence of
an $\approx 0.1$ dip near 0.5\,Hz (Observation 5) and near 2\,Hz (all other
observations).  The evidence for this dip is weak in Observation 5
(comparing band A to D, the 0.45\,Hz point is 2$\sigma$ below unity, and
the adjacent 0.64\,Hz point is only 1$\sigma$ below unity), but is somewhat
stronger for Observation 7 and the other observations (comparing band A to
D for Observation 7, the 1.8\,Hz point is $> 3\sigma$ below unity, and the
adjacent 2.5\,Hz point is $\approx 2.7\sigma$ below unity).  These dip
frequencies, however, are approximately the frequencies at which the two
broad Lorentzian and power law fit components (cf. Figure~\ref{fig:psdfits}
and Tab.~\ref{tab:tqpo}) overlap.  Thus we hypothesize that there are
indeed multiple broad-band processes occuring in GX~339--4 that are
individually coherent but that are incoherent with one another.  As with
Cygnus~X--1, we further hypothesize that the loss of coherence at
high-frequency is indicative of mulitple incoherent processes, possibly
`flares' feeding the corona on \emph{dynamical} timescales
(\cite{nowak:98a,nowak:98b}).

\begin{figure*}
\centerline{
\psfig{figure=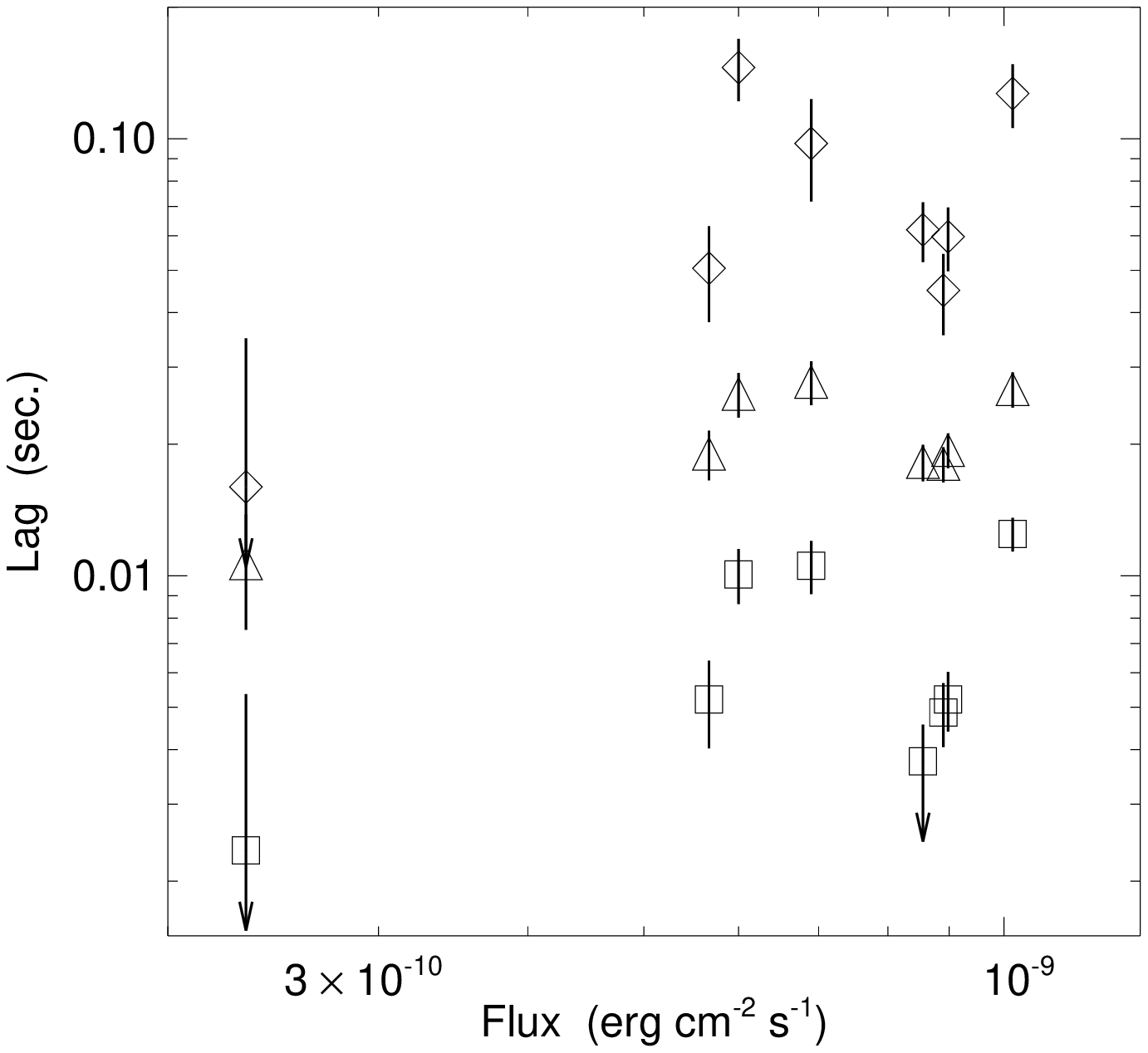,width=0.45\textwidth} 
}
\bigskip
\caption{\small Time lags at Fourier frequencies 0.1\,Hz (diamonds), 0.9\,Hz
  (triangles), and 2.5\,Hz (squares) as a function of measured 3--9\,keV
  flux. \label{fig:fluxlag}}
\end{figure*}

In Figure~\ref{fig:cross} we also show the energy-dependent and Fourier
frequency-dependent time delays.  The overall structure is very similar to
that observed in the hard state of Cyg~X--1
(\cite{miyamoto:89a,miyamoto:92a,crary:98a,nowak:98a}).  In the regions not
dominated by noise statistics, the hard photon variability always lags the
soft photon variability, and the delays decrease with increasing Fourier
frequency.  The detailed frequency-dependent structure of the delays,
however, is more complicated than a simple power law.  For example, the
time lags between bands A and D for Observation 7 show a flattened region
near 1\,Hz.  As for Cyg~X-1, there is a large dynamic range in the time
delays, with the longest time delays ($\approx 0.1$\,s) being much longer
than typical characteristic timescales of a small corona (cf.
\cite{nowak:98b}).  Also as has been observed for Cyg~X--1
(\cite{miyamoto:89a,miyamoto:92a,crary:98a,nowak:98a}), the time delay
observed in GX~339$-$4 is consistent with scaling as the logarithm of the
ratio of the two energies being compared. This latter fact has prompted
\citey{kazanas:97a} (hereafter KHT) to suggest that the time-delays are
related to photon propagation timescales in a very large ($R \approx
10^5~GM/c^2$) corona.
In Figure~\ref{fig:fluxlag} we show the measured time delay for three
Fourier frequencies (0.1\,Hz, 0.9\,Hz, 2.5\,Hz) as a function of the
measured 3--9\,keV flux. The lowest flux observation consistently shows the
shortest time delays at nearly all Fourier frequencies. The highest flux
observation shows the longest time delays at many Fourier frequencies, and
shows at least the second longest time delays at nearly all Fourier
frequencies. Observations at intermediate fluxes are scattered both
positively and negatively about an extrapolation between the low- and
high-flux point.  It has been previously shown that the time delay
decreases in Cyg~X-1 as it transits from the hard to soft state (cf.
\cite{cui:97b}). Here we present a possible correlation between the
magnitude of the time delay and the energy flux for the hard state within a
single source.  The strong decrease of the time delay for the lowest flux
observation is counter to the simplest expectations if the coronal size
increases with decreasing flux (e.g. \cite{esin:97c}), or if characteristic
`propagation speeds' in the corona decrease with decreasing luminosity
(\cite{nowak:98b}). However, if the coronal size is decreasing with
decreasing luminosity then this observed decrease is understandable in
terms of propagation models, whether it be propagation of photons (KHT) or
propagation of some sort of other disturbance (\cite{nowak:98b}).

\begin{figure*}
\centerline{
\psfig{figure=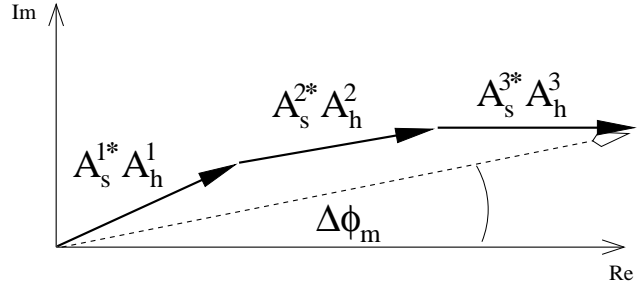,width=0.45\textwidth,angle=270} 
}
\bigskip
\caption{\small Vector analogy for phase lags and coherence. Cross power
  spectra (${A_s^1}^*{A_h^1}$, ${A_s^2}^*{A_h^2}$, ${A_s^3}^*{A_h^3}$) can
  be considered as vectors in the complex plane.  The observed phase lag,
  $\Delta \phi_m$, is the angle of the vector sum of the individual
  components of the cross power spectrum.  The coherence function is the
  square of the ratio of the magnitude of this sum compared to the
  magnitude of the sum if all the vectors had the same direction.}
\label{fig:complex}
\end{figure*}

We have found one potential correlation between the measured time lags and
the coherence function, related to the vector analogy for the phase lags
and coherence function discussed by \citey{nowak:98a}. If we take the
Fourier transform of a soft X-ray lightcurve, $A_s(f)$, and a Fourier
transform for the hard X-ray lightcurve, $A_h(f)$, the cross spectrum is
given by $A_s^*(f) A_h(f)$, which can be considered as a vector in the
complex plane (cf. Fig.~\ref{fig:complex}).  As discussed by
\citey{vaughan:97a}, the magnitude and angle (corresponding to the phase
delay between hard and soft variability) of this vector is well-defined if
the coherence function is unity.

If the net observed cross spectrum, however, is made up of the sum of
individually coherent processes, it is possible that the net observed
coherence will be \emph{\it less} than unity.  As was noted by
\citey{vaughan:97a}, specifically eq.~(10), a sum of individually coherent
processes is itself coherent if and only if each process has the same
identical transfer function from soft to hard photon variability.  In terms
of the vector analogy, this is a statement that the vectors representing
each process within the sum all lie in the same direction.  The coherence
function in such a case is the square of the ratio of the magnitude of this
sum compared to the magnitude of the sum if all the vectors had the same
angle.

Let us consider the following special case of eq.~(10) from
\citey{vaughan:97a}. Assume that there are multiple input (soft)
processes, indexed by $i$, with Fourier amplitudes $A_s^i (f)$.  Let
us assume that \emph{each} of these input processes leads to an output
process with Fourier amplitude $A_h^i (f)$ that has a \emph{coherent}
phase delay of $\Delta \phi_i (f)$.  (Here we take the Fourier
amplitudes to be real quantities.) Finally, assume that the input
processes are incoherent with one another, and likewise that the
output processes are incoherent with one another. Generalizing
eq.~(10) of \citey{vaughan:97a}, the resulting measured coherence
function will then be
\begin{eqnarray}
\gamma_m^2(f) &\approx& \left ( \sum_i {A_s^i}^2 {A_h^i}^2 + 
     \sum_{i \ne j} {A_s^i}{A_h^i} {A_s^j}{A_h^j} \left 
     [ 1 - \frac{( \Delta \phi_j - \Delta \phi_i )^2} {2} \right ] 
     \right ) 
\cr
   && ~\times~ \left ( \sum_i {A_s^i}^2 \sum_j {A_h^i}^2 \right )^{-1} 
\label{eq:coh}
\end{eqnarray}
where we have adopted a small angle approximation.  Under the same
assumptions, the net measured phase lag will be given by
\begin{equation}
\Delta \phi_m (f) \approx \frac{\sum_i A_s^i A_h^i ~\Delta \phi_i}
     {\sum_i A_s^i A_h^i} ~~.
\label{eq:lagfit}
\end{equation}
That is, the measured phase lag is simply the weighted mean of the
individual phase lags. We illustrate this situation in
Fig.~\ref{fig:complex}. For such a model as this, the net observed phase
delay is related to the net observed coherence function, and both in turn
are related to the observed soft and hard PSDs. Fit parameters are the
amplitudes ($A_s^i$, $A_h^i$) of the individual components of the soft
and hard PSDs, and the phase lags ($\Delta \phi_i$) between the soft and
hard variability for each of these components.

\begin{figure*}
\centerline{
\psfig{figure=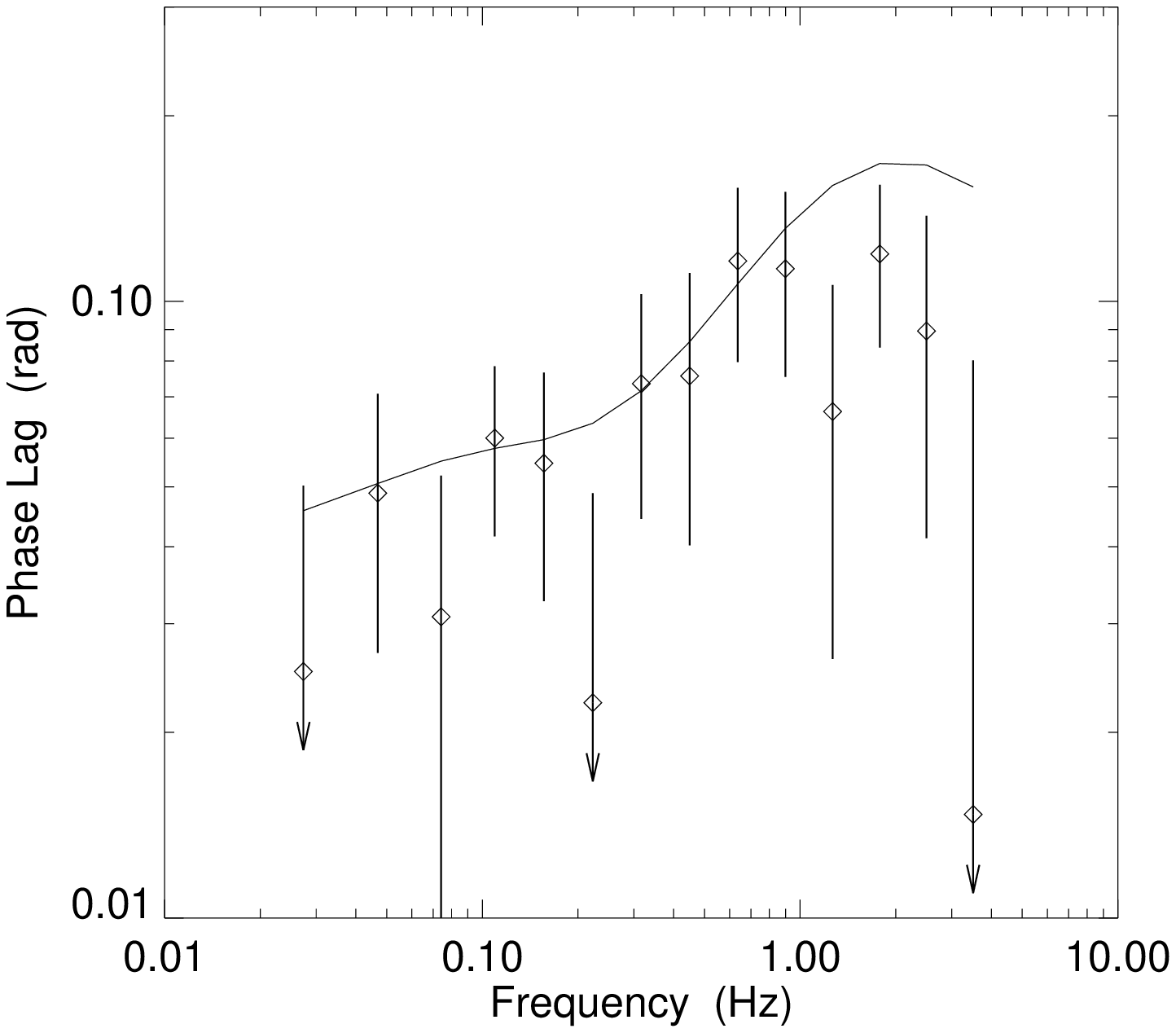,width=0.45\textwidth} ~
\psfig{figure=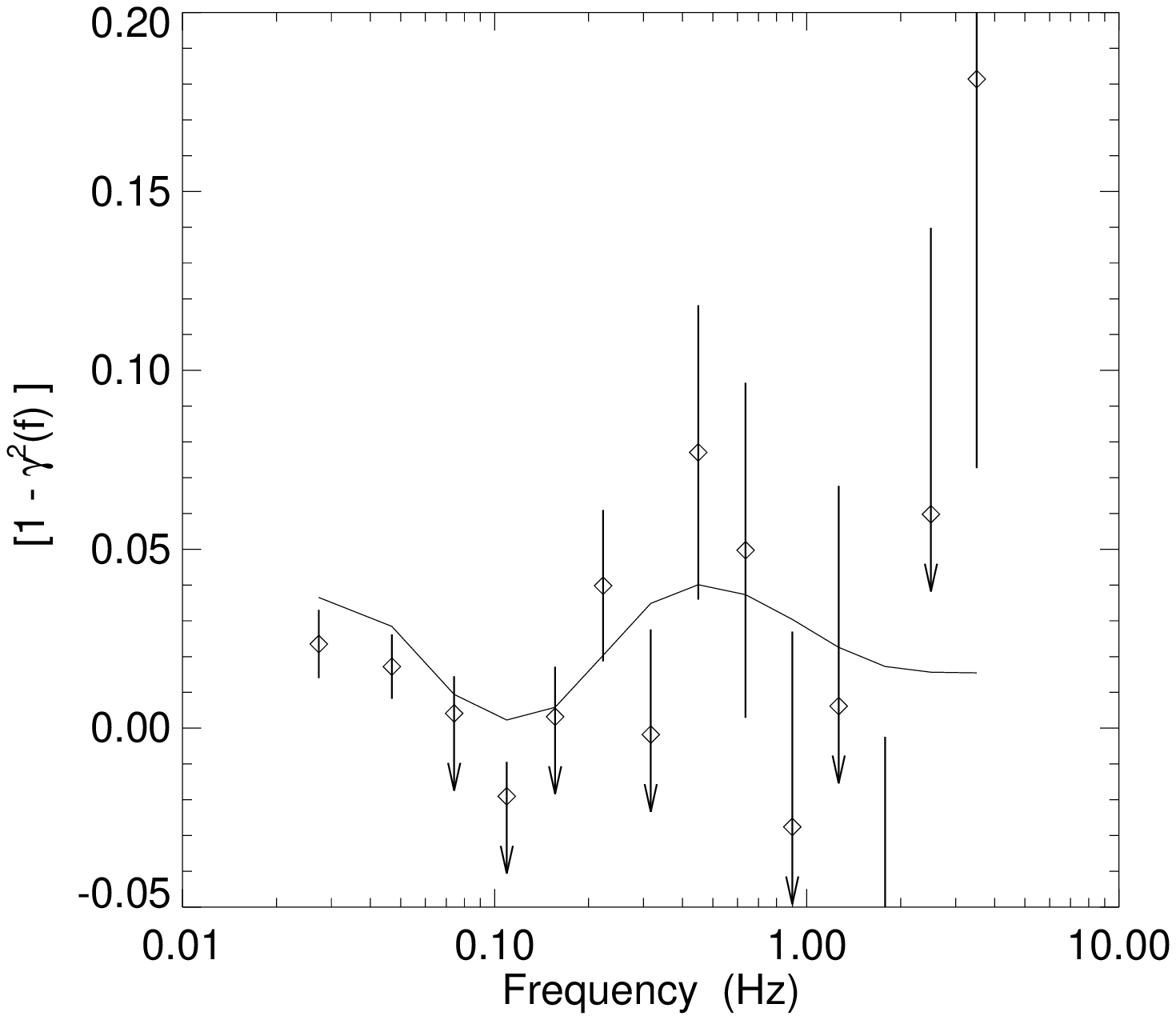,width=0.45\textwidth}
}
\centerline{
\psfig{figure=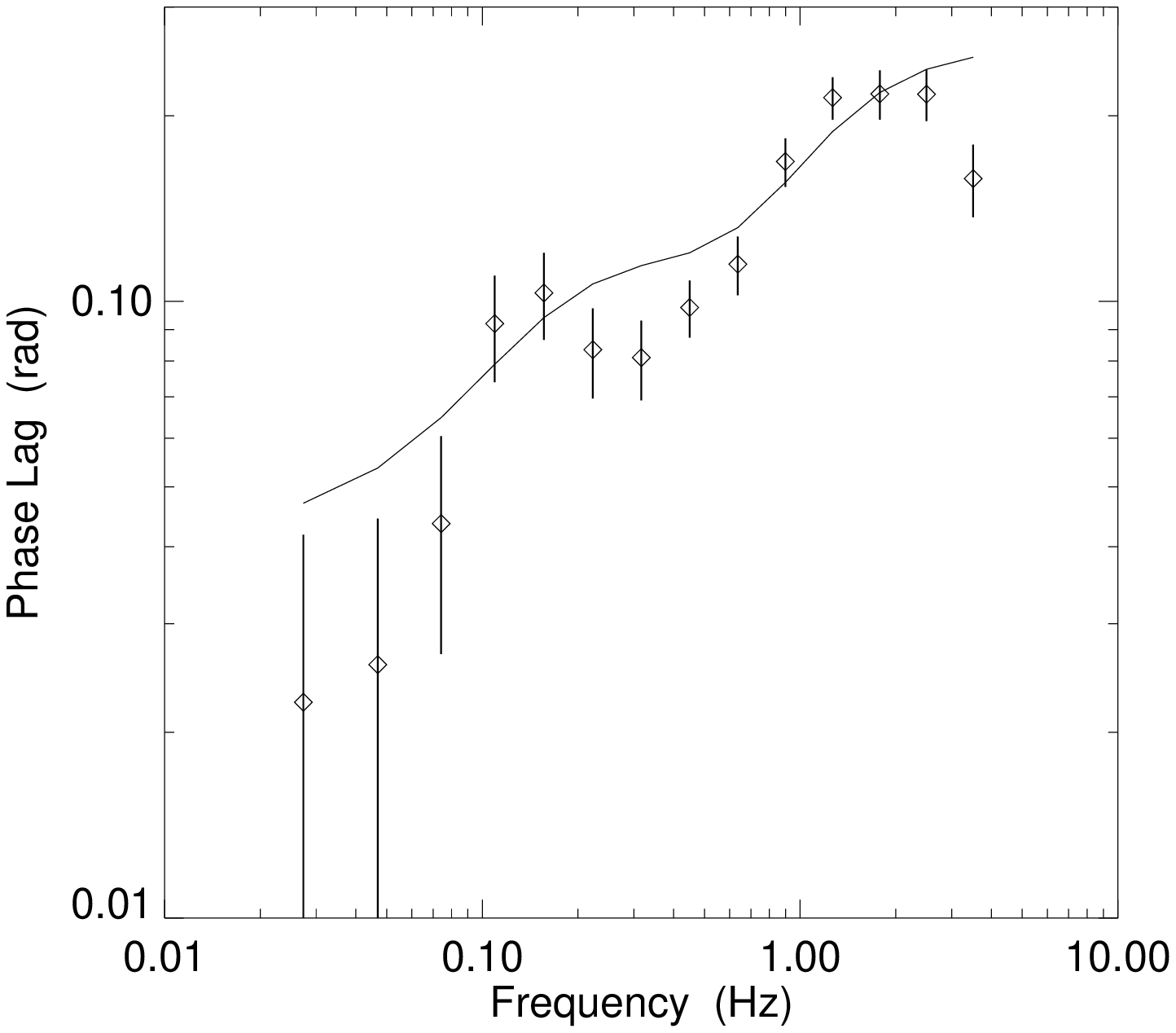,width=0.45\textwidth} ~
\psfig{figure=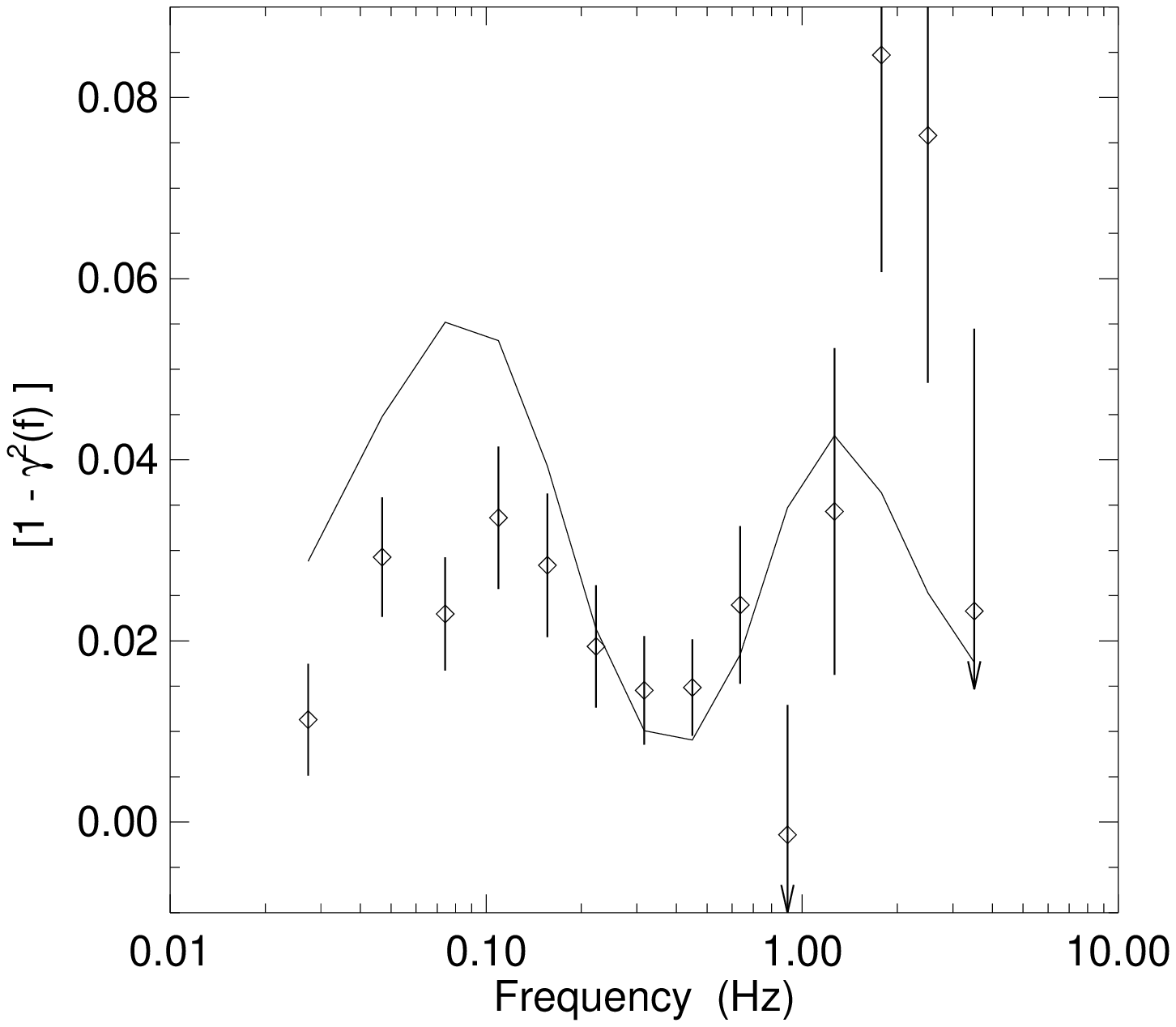,width=0.45\textwidth}
}
\bigskip
\caption{\small {\it Left:} Phase lags between energy band A and D as a
  function of Fourier frequency for Observation 5 (top) and Observation 7
  (bottom). (All points are hard variability lagging soft variability.)
  {\it Right:} $[1 - \gamma_m^2(f)]$, where $\gamma_m^2(f)$ is the
  measured, noise subtracted coherence function between energy band A and D
  for Observation 5 (top) and Observation 7 (bottom).  All data has been
  logarithmically binned over frequencies $f \rightarrow 1.4~f$. Solid
  lines are the best fit results for fitting equations of the form of
  eq.~(2) and eq.~(3) (see text and Table~2 for further explanation).}
\label{fig:lagfit}
\end{figure*}

We have searched for such a connection between the Fourier phase lag
and the coherence function by applying equations~(\ref{eq:coh}) and
(\ref{eq:lagfit}) to the data.  We have assumed that there are three
components to the PSD: a low and high frequency Lorentzian and a power
law, as for the fits presented in Table~\ref{tab:tqpo}. In the fitting
procedure we assumed that all three $\Delta \phi_i (f)$ were
independent of Fourier frequency.  We refit the PSD data
simultaneously while fitting the phase lag and coherence data.  We
searched for the minimum of the sum of the $\chi^2$ from the four data
sets being fit: soft X-ray PSD, hard X-ray PSD, phase lags, and
coherence.  We fit the PSDs over the same frequency range and with the
same frequency binning as in Table~\ref{tab:tqpo}; however, we only
fit the coherence and phase lags over the range $f=0.2$--4\,Hz.  This
was the frequency range over which the phase lags were least affected
by Poisson noise, \emph{and} it avoided the strong loss of coherence
at high frequency.  These strong high frequency coherence losses, as
we discuss further below, may be due to nonlinear processes, rather
than be due to the sum of linear processes (cf.  \cite{vaughan:97a}).
The results for these fits are presented in Figure~\ref{fig:lagfit}
and in Table~\ref{tab:tcoh}.

\begin{table*}
\caption{\small Best fit Fourier phase lags between energy channels A and
  D, where we have used equations of the form of eq.~(2) and eq.~(3) and
  where we have assumed three separate components to the soft and hard
  X-ray PSD.  PSD components are as in Table~1: a low frequency Lorentzian
  ($l1$), a power law ($pl$), and a high frequency Lorentzian ($l2$).  PSD
  data, logarithmically binned over frequencies $f \rightarrow  1.1~f$, were
  included in the fitting process with the resulting fit parameters always
  being within the error bars shown in Table~1. Phases--- $\Delta \phi_{l1}$,
  $\Delta \phi_{pl}$, $\Delta \phi_{l2}$--- are the Fourier phase lags
  between hard and soft variability for each PSD fit component
  individually, and positive phase corresponds to hard variability lagging
  behind soft variability. The phase lags were assumed to be constant as a
  function of Fourier frequency.}\label{tab:tcoh}
\medskip
{\small
\input{tcoh} }
\end{table*}

Assuming that the PSD, phase lags, and coherence are the result of
summing three independent processes (a power-law plus two
Lorentzians), we see that equations~(\ref{eq:coh}) and
(\ref{eq:lagfit}) seem to provide a rough description of the time lag
and coherence data.  The fits make plausible that there is indeed a
deeper underlying connection between the time lag and coherence data.
One feature of these fits is notable.  Namely, in order to produce
coherence drops as large as are seen, one needs to add linear
processes with greatly varying intrinsic time lags. In fact, one
process, the power law, has nearly no time delay between soft and hard
variability, whereas the high frequency Lorentzian process is seen to
require even intrinsically longer time lags than the already very long
time lags that are measured.

\section{Discussion}\label{sec:discuss}

Let us consider these results in light of two models: the Comptonization
model of KHT and `shot noise' models (cf.,
\cite{terrell:72a,sutherland:78a,priedhorsky:79a,miyamoto:89a,lochner:91a,nowak:94a},
hereafter N94; \cite{belloni:97a,poutanen:98a}, hereafter PF; and
references therein).  In the former model, a source of white noise is
assumed to be present at the center of the system, which is then
``filtered'' by passage through a Compton corona.  The PSD is attenuated on
those timescales that are \emph{shorter} than the characteristic diffusion
timescales through the corona.  Therefore, the PSD is more attenuated at
higher frequencies than at lower frequencies, which leads to the
characteristic power law PSDs that are observed in hard state BHC.  Time
delays are created by the difference in diffusion times through the corona
for hard and soft photons.  Photons that scatter over large radii will have
their intrinsic high frequency variability wiped out; therefore, any
observed high frequency variability must be due to photons that scattered
over short path lengths.  High frequency variability thus exhibits short
time delays between hard and soft photon variability.  Low frequency
variability potentially can be observed from photons that have scattered
over large path lengths, and thus it can exhibit longer time delays between
hard and soft variability.  Extremely large coronal sizes of ${\cal
  O}(10^5~GM/c^2)$ are required to produce the longest observed time lags.

In shot noise models the lightcurve is assumed to be composed of
statistically distributed shots of (possibly) varying profiles (cf.
\cite{lochner:91a} for detailed references and applications to Cyg~X-1).
Shot and distribution parameters are fit to various moments and statistics
of the observed lightcurves.  Time delays between hard and soft variability
are assumed to be due to differing shot profiles and/or shot distributions
in different energy bands (cf.  \cite{miyamoto:89a}; N94; PF).  Recently,
PF have associated the energy dependent shots with the `energization' of a
corona, which was parameterized by a series of equilibrium models where the
energy input to the corona was a function of time.

Does the data prefer one model over the other?  The detailed structure of
the PSD is likely to be more difficult to produce in the KHT model.
Although the observed PSD can be modeled as a singly broken power law with
an `absorption' feature, it is difficult to imagine a Compton corona
configuration that would act as such a `notch filter'.  It is somewhat
easier to imagine, however, two or more preferred shot durations in the
distribution of shot timescales. A suitable distribution of shot timescales
could easily reproduce the type of PSD fits presented in \S\ref{sec:psd}.

Can the flux dependence of the PSD be reproduced? This depends upon whether
the coronal radius decreases or increases as the observed flux decreases.  The
former possibility can be consistent with our `sphere+disk' corona fits to
the spectral data (paper I).  The latter is more consistent with the ADAF
models to the spectral data (paper I), as lower luminosity often implies a
larger `transition radius' to advective flow.  The transition radius in the
ADAF models is the radius at which the flow transits from being
geometrically thin, optically thick, and radiatively efficient to being
geometrically thick, optically thin, and radiatively inefficient.  The
larger this transition radius, the smaller the overall accretion efficiency
(cf.  \cite{narayan:96e,esin:97c}). Lower luminosities can be associated
with lower overall efficiencies.  The dependence of the coronal radius on
observed X-ray luminosity is less constrained for our `sphere+disk' coronal
models (\cite{dove:97b}; paper I).  Larger coronae, however, can produce
larger observed X-ray fluxes in these models.

In the KHT model, the larger the corona the more attenuated the high
frequency PSD will be.  This will cause the PSD to apparently `shift'
towards lower frequency.  Similarly, shot noise models usually associate
the shot timescales with characteristic accretion flow timescales.  What
sets these timescales are usually not explicit in shot noise models;
however, one expects the timescales to \emph{increase} for larger coronae,
consistent with the expectations of the ADAF models.  That is, we expect to
see the lower characteristic frequencies associated with the outer
accretion flow.  Associating a smaller corona with lower luminosity, as can
be fitted within the context of `sphere+disk' corona models, would lead to
trends opposite of the observations.

What is the expected relation between the PSD and the observed time
lags for these models?  If the coronal size decreases with decreasing
luminosity, then the KHT model agrees with the observations.  The
shorter scattering path lengths of a small corona will produce
characteristically shorter lags between the hard and soft photon
variability.  The KHT model also correctly reproduces the observed
logarithmic energy dependence of the time lags.  However, it is
difficult for the KHT model to reconcile the lower frequency PSDs
\emph{simultaneously} with shorter time lags.  Predictions for shot
noise models are more dependent upon the specifics of each model.

Energy dependent shots were first suggested by \citey{miyamoto:89a} as
the cause of the hard X-ray variability lags observed in Cyg~X--1.
This model contained eight input shot profiles (four shot durations in
two energy bands), and did not relate the timescales to specific disk
physics.  The magnitude and frequency dependence of the hard X-ray
lags are adjustable by changing the duration of the hard shots
relative to the soft shots.  In describing observations of the `very
high state' timing properties of GX~339--4, N94 related the shot
durations to viscous and thermal instability timescales in an
accretion disk.  The shots considered in this work were also
temperature dependent and became hotter as the shots progressed.  This
model reproduced the observed energy dependence of the lags, with the
exception that it did not reproduce the 1.2--2.3\,keV photon
variability lagging behind the 2.3--4.6\,keV photon variability.  As
discussed in N94 and \citey{nowak:97a}, any shot model where the
characteristic temperature or hardness of the shot smoothly increases
as the shot rises will reproduce time lags qualitatively similar to
those observed in Cyg~X-1 and GX~339-4. In the model of PF, the shot
timescales are not related to any specific accretion timescales;
however, the time-dependent shot hardness is related to a series of
equilibrium Compton coronae models.  There is a one-to-one
correspondance between the shot amplitude and its hardness in this
model.  The PF model correctly reproduces the magnitude and energy
dependence of the observed time lags.

We expect the following qualitative behavior for the models of both N94 and
PF. If we scale the shot duration to longer timescales, thereby shifting
the PSD to lower frequencies, we also expect the timelags to
\emph{increase}, which is \emph{contrary} to the observations.  The only
way to have a shot produce a lower frequency PSD \emph{and} lead to shorter
time lags is to alter the spectrum of the shot as well as lengthen its
duration.  Whereas we do see changes in flux from GX~339--4, we do not see
large changes in the best fit spectral parameters such as optical depth or
coronal temperature.  The observed energy spectra do not leave much room
for varying the spectra of the presumed shots in order to fit both the high
frequency PSD/long time lag data and the low frequency PSD/short time lag
data that are observed. (It is possible, however, for the relative
contributions to the spectra from a `steady component' and a `shot
component' vary in such a way that the average spectrum remains relatively
unchanged.  The rms amplitude of the PSD significantly increases for the
faintest observation, and therefore its average spectrum \emph{does} have a
greater contribution from the shots.)

What is the expected coherence function for these various models? Both
the KHT model and the PF model produce unity coherence, whereas the
model of N94 produces coherence substantially less than unity
(\cite{nowak:97a}).  The KHT model only considers static coronae,
whereas the PF model uses the same exact linear Comptonization
spectrum for each individual shot. We expect both of these situations
to produce unity coherence.  The N94 model assumes both a sum of
different shot spectra \emph{and} each shot represents a
\emph{nonlinear} transfer function from soft to hard variability
(\cite{nowak:97a,vaughan:97a}).

Similar to Cyg~X--1, GX~339--4 was seen to have near unity coherence
over a broad range of frequencies, with a rolloff at $\approx
3$--10\,Hz.  Unlike Cyg~X-1 (\cite{nowak:98a}), however, there was no
evidence for a loss of coherence below $\approx 0.02$\,Hz, but there
was evidence for a coherence dip near 2\,Hz (0.5\,Hz for Observation
5).  At first glance this seems consistent with both the KHT and PF
models.  We note, however, that the coherence function dips to as low
as 0.9.  Although still a large coherence value, a dip this low is
very difficult to produce by summing different \emph{linear} transfer
functions with different phase delays, \emph{if} all the hard-to-soft
variability phase differences for the individual transfer functions
are approximately the same magnitude as the \emph{observed} phase
lags.  On the other hand, the nonlinear transfer functions of the N94
model are seen to lead to far stronger losses of coherence, with
typical coherence values being $\approx 0.1$--0.3
(\cite{nowak:97a}). Such coherence values are characteristic of the
coherence function observed at high frequency.  (We previously have
suggested that the observed loss of coherence above $\approx
3$--10\,Hz may be related to nonlinear effects due to flaring activity
feeding a corona on \emph{dynamical} timescales; \cite{nowak:98a}.)

The question put forth in \citey{vaughan:97a} was: why are observed
coherences so close to unity?  With the observations presented here,
and in light of the discussion of \S\ref{sec:lag} and
equations~(\ref{eq:coh}) and (\ref{eq:lagfit}), we might wish to
modify this question to: when the coherence function is so close to
unity, why is it as \emph{low} as observed?  We argued in
\S\ref{sec:lag} that if there truly is a relationship between the time
lags and the coherence, then there must be a mix of processes with a
\emph{broad} range of time lags.  The fits to the phase lag and
coherence data suggested one component with near zero lag between hard
and soft variability. The fits further suggested that the higher
frequency Lorentzian component exhibited characteristically longer
time lags (at a fixed Fourier frequency) than the lower frequency
Lorentzian component.  This possibility was not considered in the
models of N94, KHT, or PF.  However, as the KHT model postulates only
one `transfer function' from soft variability to hard variability
(i.e., diffusion through the corona), it is more difficult to
reconcile this model with the conjecture that phase lags and coherence
are created by a sum of independent processes.  Shot noise models,
however, can introduce multiple, albeit currently unknown, independent
processes in the manner suggested by the fits to the data.

\section{Summary}\label{sec:summ}

We have presented timing analysis of a series of 8 RXTE observations
of the black hole candidate GX~339--4.  On long timescales, there is
evidence of a 240 day periodicity in the ASM lightcurve.  This is not
a strictly periodic feature, but is probably more like a
`characteristic timescale'.  Such a timescale is consistent with
warping and precession timescales (\cite{pringle:96a,maloney:96a});
however, as we discuss in paper I, the evidence points more towards
coronal size variations than to inclination effects.

In terms of characterizing the short timescale variability, we see
that the observations break up into two sets.  The seven brightest
observations span a range of two in observed 3--9\,keV flux and all
have comparable timing properties.  These observations all show
evidence of an $\approx 0.3$\,Hz QPO.  The properties of this QPO are
not strictly steady; however, there are no obvious correlations with
the flux of the source.

The faintest observation, which is 5 times fainter than the brightest
observation, had: larger amplitude variability, characteristic PSD
frequencies that were a factor of three lower than observed for the
other PSDs, and showed shorter time lags between hard and soft
variability.  This latter feature, albeit with a great deal of scatter
about the trend, was also mirrored in the flux dependence of the time
lags observed for the brightest observations.

All PSDs were reasonably well-fit by the sum of three fit components:
a power law, a low frequency Lorentzian, and a high frequency
Lorentzian.  We further suggested that the observed Fourier
frequency-dependent phase lags and coherence could be explained by
summing these three components with differing intrinsic time delays
between hard and soft photon variability.

We discussed all these possibilities in light of various theoretical
models.  The short time delays for the lowest flux observation appear to be
more in agreement with `propagation models' (KHT; \cite{nowak:98b})
\emph{if the coronal size decreases with decreasing flux}. If, however, as
suggested by ADAF models the coronal (i.e., advection dominated) region
grows with decreasing flux, the trends observed in the time delays are
\emph{counter} to the theory.  Conversely, the low flux/low frequency PSD
observations are more in agreement with the models where the coronal region
grows with decreasing luminosity.  Shot noise models are more likely able
to be adapted to explain simultaneously the phase lags and coherence as a
sum of independent linear components.

None of the models currently address the complicated nonlinear
processes that may be occurring on \emph{dynamical} timescales.  Such
processes, which could be the `flares' that energize the corona and/or
the observed radio-emitting outflow
(\cite{fender:97a,corbel:98a,hanni:98a}; paper I), could be the cause
of the strong loss of coherence seen at high frequency in both
GX~339--4 and Cyg~X--1 (\cite{nowak:98a}), as well as the cause of the
`flattening' with increasing photon energy observed at the
high-frequency end of the Cyg~X--1 PSD (\cite{nowak:98a}).

\acknowledgements We would like to acknowledge useful conversations with T.
Aldcroft, M.  Begelman, R. Fender, P. Maloney, T. DiMatteo, K. Pottschmidt,
R. Staubert, and B.  Vaughan. This work has been financed by NASA Grants
{NAG5-3225}, {NAG5-4731}, and by a travel grant to J.W. from the DAAD.


\end{document}

%% file: tqpo_pow.tex
\hfill\hbox{
\begin{tabular}{lllllllllll}
\hline
\hline
Obs. & Band & R$_1$ & $f_{1}$ & $Q_1$ & A & $\Gamma$  & R$_2$ & $f_2$ & $Q_2$ & $\chi^2$ \\ 
&& ($\times 10^{-2}$) & (Hz) & & ($\times 10^{-3}$) & & ($\times 10^{-2}$) & (Hz) \\
\hline
\noalign{\vspace*{0.5mm}}
1 & A &  \errtwo{24.}{ 2.}{ 2.} & \errtwo{ 0.33}{ 0.04}{ 0.04} & \errtwo{ 1.2}{ 0.3}{ 0.2} & \errtwo{  7.5}{  1.4}{  1.7} & \errtwo{-1.23}{ 0.06}{ 0.11} & \errtwo{19.}{ 3.}{ 2.} & \errtwo{ 2.2}{ 0.3}{ 0.6} & \errtwo{ 0.9}{ 0.2}{ 0.3} &  102. \\
\noalign{\vspace*{0.5mm}} 
1 & B & \errtwo{26.}{ 2.}{ 2.} & \errtwo{ 0.31}{ 0.04}{ 0.03} & \errtwo{ 1.0}{ 0.3}{ 0.2} & \errtwo{  5.4}{  1.4}{  1.8} & \errtwo{-1.28}{ 0.07}{ 0.18} & \errtwo{21.}{ 3.}{ 2.} & \errtwo{ 2.6}{ 0.3}{ 0.6} & \errtwo{ 0.8}{ 0.2}{ 0.3} &  116. \\
\noalign{\vspace*{0.5mm}} 
1 & C & \errtwo{25.}{ 2.}{ 2.} & \errtwo{ 0.31}{ 0.04}{ 0.03} & \errtwo{ 1.0}{ 0.3}{ 0.2} & \errtwo{  4.7}{  1.6}{  2.5} & \errtwo{-1.26}{ 0.08}{ 0.32} & \errtwo{21.}{ 4.}{ 2.} & \errtwo{ 2.4}{ 0.4}{ 0.8} & \errtwo{ 0.7}{ 0.2}{ 0.3} &   84. \\
\noalign{\vspace*{0.5mm}} 
1 & D & \errtwo{20.}{ 1.}{ 2.} & \errtwo{ 0.35}{ 0.03}{ 0.02} & \errtwo{ 1.3}{ 0.3}{ 0.2} & \errtwo{  5.6}{  1.0}{  1.1} & \errtwo{-1.17}{ 0.04}{ 0.09} & \errtwo{16.}{ 3.}{ 2.} & \errtwo{ 3.2}{ 0.3}{ 0.6} & \errtwo{ 1.1}{ 0.2}{ 0.4} &  112. \\
\noalign{\vspace*{0.5mm}} 
1 & T & \errtwo{25.}{ 2.}{ 2.} & \errtwo{ 0.30}{ 0.04}{ 0.03} & \errtwo{ 0.9}{ 0.3}{ 0.1} & \errtwo{  5.7}{  0.9}{  0.9} & \errtwo{-1.19}{ 0.04}{ 0.06} & \errtwo{18.}{ 1.}{ 1.} & \errtwo{ 2.9}{ 0.1}{ 0.3} & \errtwo{ 1.1}{ 0.1}{ 0.2} &  129. \\
\noalign{\vspace*{1.5mm}} 
2 & A & \errtwo{27.}{ 3.}{ 4.} & \errtwo{ 0.28}{ 0.06}{ 0.03} & \errtwo{ 0.9}{ 0.4}{ 0.1} & \errtwo{  4.6}{  2.7}{  3.4} & \errtwo{-1.46}{ 0.13}{ 0.51} & \errtwo{23.}{ 4.}{ 3.} & \errtwo{ 2.2}{ 0.4}{ 0.7} & \errtwo{ 0.8}{ 0.3}{ 0.2} &  103. \\
\noalign{\vspace*{0.5mm}} 
2 & B & \errtwo{26.}{ 2.}{ 3.} & \errtwo{ 0.29}{ 0.05}{ 0.03} & \errtwo{ 0.9}{ 0.3}{ 0.1} & \errtwo{  6.3}{  1.8}{  1.7} & \errtwo{-1.30}{ 0.08}{ 0.15} & \errtwo{20.}{ 3.}{ 2.} & \errtwo{ 2.6}{ 0.3}{ 0.5} & \errtwo{ 0.9}{ 0.2}{ 0.3} &   73. \\
\noalign{\vspace*{0.5mm}} 
2 & C & \errtwo{22.}{ 2.}{ 3.} & \errtwo{ 0.32}{ 0.04}{ 0.03} & \errtwo{ 1.3}{ 0.4}{ 0.2} & \errtwo{  6.7}{  1.8}{  2.9} & \errtwo{-1.27}{ 0.08}{ 0.23} & \errtwo{21.}{ 4.}{ 3.} & \errtwo{ 2.2}{ 0.5}{ 1.0} & \errtwo{ 0.7}{ 0.3}{ 0.4} &   74. \\
\noalign{\vspace*{0.5mm}} 
2 & D & \errtwo{21.}{3.}{3.} & \errtwo{0.31}{0.04}{0.03} & \errtwo{1.1}{0.4}{0.2} & \errtwo{3.7}{2.0}{3.4} & \errtwo{-1.35}{0.13}{0.62} & \errtwo{20.}{3.}{3.} & \errtwo{2.2}{0.4}{0.7} & \errtwo{0.6}{0.3}{0.3} & 94. \\
\noalign{\vspace*{0.5mm}} 
2 & T & \errtwo{25.}{ 2.}{ 3.} & \errtwo{ 0.28}{ 0.05}{ 0.03} & \errtwo{ 0.9}{ 0.3}{ 0.1} & \errtwo{  6.1}{  1.1}{  0.9} & \errtwo{-1.24}{ 0.04}{ 0.06} & \errtwo{18.}{ 1.}{ 1.} & \errtwo{ 2.8}{ 0.1}{ 0.3} & \errtwo{ 1.0}{ 0.1}{ 0.2} &  102. \\
\noalign{\vspace*{1.5mm}} 
3 & A & \errtwo{24.}{ 3.}{ 3.} & \errtwo{ 0.32}{ 0.03}{ 0.03} & \errtwo{ 1.3}{ 0.4}{ 0.2} & \errtwo{  4.6}{  2.8}{  2.9} & \errtwo{-1.46}{ 0.16}{ 0.51} & \errtwo{24.}{ 3.}{ 4.} & \errtwo{ 1.8}{ 0.5}{ 0.3} & \errtwo{ 0.6}{ 0.3}{ 0.1} &   77. \\
\noalign{\vspace*{0.5mm}} 
3 & B & \errtwo{25.}{ 2.}{ 2.} & \errtwo{ 0.30}{ 0.04}{ 0.03} & \errtwo{ 1.1}{ 0.3}{ 0.2} & \errtwo{  5.3}{  1.8}{  1.6} & \errtwo{-1.32}{ 0.09}{ 0.16} & \errtwo{22.}{ 3.}{ 2.} & \errtwo{ 2.2}{ 0.3}{ 0.4} & \errtwo{ 0.8}{ 0.2}{ 0.2} &   86. \\
\noalign{\vspace*{0.5mm}} 
3 & C & \errtwo{22.}{2.}{2.} & \errtwo{0.32}{0.03}{0.02} & \errtwo{1.3}{0.4}{0.2} & \errtwo{5.9}{1.7}{3.4} & \errtwo{-1.27}{0.06}{0.38} & \errtwo{22.}{3.}{2.} & \errtwo{2.0}{0.4}{0.4} & \errtwo{0.6}{0.3}{0.3} & 73. \\
\noalign{\vspace*{0.5mm}} 
3 & D &  \errtwo{19.}{1.}{2.} & \errtwo{0.35}{0.02}{0.02} & \errtwo{1.5}{0.3}{0.2} & \errtwo{5.0}{1.3}{2.7} & \errtwo{-1.25}{0.08}{0.33} & \errtwo{19.}{3.}{2} & \errtwo{2.2}{0.4}{0.4} & \errtwo{0.7}{0.3}{0.3} & 92. \\
\noalign{\vspace*{0.5mm}} 
3 & T & \errtwo{24.}{ 1.}{ 3.} & \errtwo{ 0.30}{ 0.04}{ 0.02} & \errtwo{ 1.1}{ 0.3}{ 0.1} & \errtwo{  5.4}{  1.0}{  0.8} & \errtwo{-1.26}{ 0.03}{ 0.08} & \errtwo{20.}{ 2.}{ 1.} & \errtwo{ 2.5}{ 0.1}{ 0.4} & \errtwo{ 0.9}{ 0.1}{ 0.2} &  119. \\
\noalign{\vspace*{1.5mm}} 
4 & A & \errtwo{19.}{ 2.}{ 3.} & \errtwo{ 0.35}{ 0.03}{ 0.03} & \errtwo{ 1.6}{ 0.7}{ 0.4} & \errtwo{ 11.5}{  1.8}{  2.1} & \errtwo{-1.17}{ 0.05}{ 0.09} & \errtwo{16.}{ 5.}{ 3.} & \errtwo{ 2.0}{ 0.5}{ 0.7} & \errtwo{ 0.9}{ 0.5}{ 0.3} &   48. \\
\noalign{\vspace*{0.5mm}} 
4 & B & \errtwo{24.}{ 2.}{ 3.} & \errtwo{ 0.31}{ 0.05}{ 0.04} & \errtwo{ 1.1}{ 0.5}{ 0.2} & \errtwo{  6.3}{  2.1}{  4.5} & \errtwo{-1.30}{ 0.09}{ 0.53} & \errtwo{20.}{ 4.}{ 3.} & \errtwo{ 2.4}{ 0.4}{ 1.1} & \errtwo{ 0.8}{ 0.3}{ 0.4} &   82. \\
\noalign{\vspace*{0.5mm}} 
4 & C & \errtwo{19.}{ 2.}{ 2.} & \errtwo{ 0.36}{ 0.03}{ 0.03} & \errtwo{ 1.5}{ 0.5}{ 0.3} & \errtwo{  7.5}{  2.1}{  5.0} & \errtwo{-1.24}{ 0.07}{ 0.47} & \errtwo{21.}{ 7.}{ 4.} & \errtwo{ 1.9}{ 0.8}{ 1.5} & \errtwo{ 0.6}{ 0.4}{ 0.5} &   62. \\
\noalign{\vspace*{0.5mm}} 
4 & D & \errtwo{17.}{ 2.}{ 2.} & \errtwo{ 0.37}{ 0.03}{ 0.03} & \errtwo{ 1.5}{ 0.5}{ 0.3} & \errtwo{  6.8}{  1.3}{  1.4} & \errtwo{-1.16}{ 0.05}{ 0.10} & \errtwo{14.}{ 4.}{ 2.} & \errtwo{ 3.0}{ 0.4}{ 0.8} & \errtwo{ 1.2}{ 0.5}{ 0.5} &   66. \\
\noalign{\vspace*{0.5mm}} 
4 & T & \errtwo{22.}{ 2.}{ 3.} & \errtwo{ 0.31}{ 0.05}{ 0.03} & \errtwo{ 1.0}{ 0.4}{ 0.2} & \errtwo{  6.5}{  1.1}{  0.9} & \errtwo{-1.23}{ 0.04}{ 0.07} & \errtwo{18.}{ 2.}{ 1.} & \errtwo{ 2.7}{ 0.2}{ 0.5} & \errtwo{ 1.0}{ 0.1}{ 0.2} &   86. \\
\noalign{\vspace*{1.5mm}} 
5 & A & \errtwo{27.}{12.}{ 8.} & \errtwo{ 0.08}{ 0.04}{ 0.05} & \errtwo{ 0.8}{ 1.0}{ 0.6} & \errtwo{  5.1}{  6.5}{  4.8} & \errtwo{-1.11}{ 1.06}{ 0.85} & \errtwo{25.}{ 7.}{10.} & \errtwo{ 1.0}{ 0.7}{ 0.7} & \errtwo{ 0.6}{ 1.1}{ 0.4} &   59. \\
\noalign{\vspace*{0.5mm}} 
5 & B & \errtwo{31.}{ 9.}{ 7.} & \errtwo{ 0.07}{ 0.04}{ 0.05} & \errtwo{ 0.6}{ 0.6}{ 0.5} & \errtwo{  3.5}{  5.0}{  3.2} & \errtwo{-1.23}{ 0.92}{ 0.73} & \errtwo{29.}{ 4.}{ 7.} & \errtwo{ 0.7}{ 0.5}{ 0.4} & \errtwo{ 0.4}{ 0.5}{ 0.2} &   46. \\
\noalign{\vspace*{0.5mm}} 
5 & C & \errtwo{35.}{ 8.}{10.} & \errtwo{ 0.04}{ 0.05}{ 0.04} & \errtwo{ 0.4}{ 0.7}{ 0.3} & \errtwo{  1.6}{  5.1}{  1.4} & \errtwo{-1.30}{ 1.10}{ 0.67} & \errtwo{28.}{ 4.}{ 5.} & \errtwo{ 1.0}{ 0.4}{ 0.6} & \errtwo{ 0.6}{ 0.3}{ 0.3} &   69. \\
\noalign{\vspace*{0.5mm}} 
5 & D & \errtwo{23.}{7.}{7.} & \errtwo{0.08}{0.04}{0.03} &
\errtwo{0.8}{0.8}{0.3} & \errtwo{5.0}{4.0}{3.8} &
\errtwo{-1.07}{0.19}{0.41} & \errtwo{20.}{7.}{8.} & \errtwo{1.1}{0.6}{0.6}
& \errtwo{0.6}{0.9}{0.4} & 49. \\ 
\noalign{\vspace*{0.5mm}} 
5 & T & \errtwo{35.}{ 5.}{ 8.} & \errtwo{ 0.03}{ 0.04}{ 0.03} & \errtwo{ 0.3}{ 0.4}{ 0.2} & \errtwo{  1.0}{  3.2}{  0.4} & \errtwo{-1.25}{ 0.60}{ 0.45} & \errtwo{27.}{ 2.}{ 4.} & \errtwo{ 0.9}{ 0.3}{ 0.3} & \errtwo{ 0.5}{ 0.2}{ 0.2} &   85. \\
\noalign{\vspace*{1.5mm}} 
6 & A & \errtwo{21.}{ 2.}{ 3.} & \errtwo{ 0.33}{ 0.04}{ 0.03} & \errtwo{ 1.4}{ 0.6}{ 0.2} & \errtwo{  9.6}{  1.7}{  1.5} & \errtwo{-1.17}{ 0.04}{ 0.09} & \errtwo{17.}{ 4.}{ 2.} & \errtwo{ 2.4}{ 0.3}{ 0.6} & \errtwo{ 1.2}{ 0.4}{ 0.5} &   83. \\
\noalign{\vspace*{0.5mm}} 
6 & B & \errtwo{25.}{ 2.}{ 3.} & \errtwo{ 0.29}{ 0.05}{ 0.03} & \errtwo{ 1.1}{ 0.4} { 0.2} & \errtwo{  6.6}{  1.7}{  1.4} & \errtwo{-1.18}{ 0.06}{ 0.10} & \errtwo{18.}{ 3.}{ 2.} & \errtwo{ 2.7}{ 0.2}{ 0.4} & \errtwo{ 1.1}{ 0.2}{ 0.3} &   89. \\
\noalign{\vspace*{0.5mm}} 
6 & C & \errtwo{22.}{ 3.}{ 3.} & \errtwo{ 0.33}{ 0.04}{ 0.02} & \errtwo{ 1.3}{ 0.5}{ 0.2} & \errtwo{  5.3}{  2.1}{  4.1} & \errtwo{-1.31}{ 0.10}{ 0.65} & \errtwo{22.}{ 4.}{ 3.} & \errtwo{ 2.0}{ 0.5}{ 1.1} & \errtwo{ 0.7}{ 0.3}{ 0.4} &   72. \\
\noalign{\vspace*{0.5mm}} 
6 & D & \errtwo{18.}{ 2.}{ 2.} & \errtwo{ 0.34}{ 0.03}{ 0.02} & \errtwo{ 1.5}{ 0.5}{ 0.2} & \errtwo{  4.7}{  1.7}{  3.1} & \errtwo{-1.26}{ 0.09}{ 0.51} & \errtwo{20.}{ 5.}{ 3.} & \errtwo{ 1.9}{ 0.6}{ 1.5} & \errtwo{ 0.6}{ 0.3}{ 0.4} &   75. \\
\noalign{\vspace*{0.5mm}} 
6 & T & \errtwo{23.}{ 1.}{ 3.} & \errtwo{ 0.30}{ 0.05}{ 0.02} & \errtwo{ 1.1}{ 0.4}{ 0.1} & \errtwo{  5.8}{  1.1}{  0.8} & \errtwo{-1.21}{ 0.04}{ 0.07} & \errtwo{19.}{ 2.}{ 1.} & \errtwo{ 2.6}{ 0.1}{ 0.3} & \errtwo{ 1.0}{ 0.1}{ 0.2} &  101. \\
\noalign{\vspace*{1.5mm}} 
7 & A & \errtwo{22.}{ 2.}{ 2.} & \errtwo{ 0.38}{ 0.03}{ 0.02} & \errtwo{ 1.7}{ 0.4}{ 0.3} & \errtwo{  4.6}{  2.5}{  2.8} & \errtwo{-1.45}{ 0.18}{ 0.37} & \errtwo{28.}{ 2.}{ 4.} & \errtwo{ 0.8}{ 0.7}{ 0.3} & \errtwo{ 0.3}{ 0.3}{ 0.1} &  115. \\
\noalign{\vspace*{0.5mm}} 
7 & B & \errtwo{24.}{ 1.}{ 2.} & \errtwo{ 0.36}{ 0.03}{ 0.02} & \errtwo{ 1.3}{ 0.4}{ 0.3} & \errtwo{  6.2}{  1.4}{  1.8} & \errtwo{-1.22}{ 0.06}{ 0.16} & \errtwo{21.}{ 2.}{ 2.} & \errtwo{ 2.1}{ 0.6}{ 0.5} & \errtwo{ 0.7}{ 0.4}{ 0.3} &  115. \\
\noalign{\vspace*{0.5mm}} 
7 & C & \errtwo{23.}{ 1.}{ 2.} & \errtwo{ 0.36}{ 0.04}{ 0.03} & \errtwo{ 1.3}{ 0.4}{ 0.2} & \errtwo{  6.1}{  1.3}{  2.1} & \errtwo{-1.15}{ 0.06}{ 0.20} & \errtwo{19.}{ 3.}{ 2.} & \errtwo{ 2.4}{ 0.5}{ 1.6} & \errtwo{ 0.7}{ 0.3}{ 0.5} &  112. \\
\noalign{\vspace*{0.5mm}} 
7 & D & \errtwo{20.}{ 1.}{ 2.} & \errtwo{ 0.39}{ 0.03}{ 0.03} & \errtwo{ 1.4}{ 0.4}{ 0.2} & \errtwo{  4.8}{  0.7}{  2.2} & \errtwo{-1.14}{ 0.04}{ 0.24} & \errtwo{16.}{ 3.}{ 1.} & \errtwo{ 2.7}{ 0.3}{ 1.5} & \errtwo{ 0.8}{ 0.2}{ 0.5} &  124. \\
\noalign{\vspace*{0.5mm}} 
7 & T & \errtwo{25.}{1.}{2.} & \errtwo{0.33}{0.04}{0.03} & \errtwo{1.0}{0.2}{0.1} & \errtwo{4.9}{0.7}{0.8} & \errtwo{-1.18}{ 0.05}{0.06} & \errtwo{19.}{ 1.}{ 1.} & \errtwo{ 2.7}{ 0.2}{ 0.3} & \errtwo{0.9}{0.1}{0.1} &  165. \\
\noalign{\vspace*{1.5mm}} 
8 & A & \errtwo{21.}{ 2.}{ 2.} & \errtwo{ 0.36}{ 0.04}{ 0.05} & \errtwo{ 1.3}{ 0.3}{ 0.3} & \errtwo{  9.7}{  1.5}{  1.8} & \errtwo{-1.20}{ 0.05}{ 0.08} & \errtwo{16.}{ 2.}{ 3.} & \errtwo{ 2.7}{ 0.3}{ 0.5} & \errtwo{ 1.2}{ 0.5}{ 0.4} &  61. \\
\noalign{\vspace*{0.5mm}} 
8 & B & \errtwo{23.}{ 5.}{ 2.} & \errtwo{ 0.36}{ 0.03}{ 0.05} & \errtwo{ 1.2}{ 0.3}{ 0.3} & \errtwo{  5.1}{  2.2}{  3.9} & \errtwo{-1.45}{ 0.15}{ 0.69} & \errtwo{23.}{ 4.}{ 4.} & \errtwo{ 1.9}{ 0.6}{ 0.7} & \errtwo{ 0.6}{ 0.3}{ 0.2} &  78. \\
\noalign{\vspace*{0.5mm}} 
8 & C & \errtwo{21.}{ 5.}{ 2.} & \errtwo{ 0.36}{ 0.03}{ 0.06} & \errtwo{ 1.2}{ 0.3}{ 0.3} & \errtwo{  5.0}{  2.4}{  3.7} & \errtwo{-1.42}{ 0.15}{ 0.71} & \errtwo{23.}{ 5.}{ 4.} & \errtwo{ 1.8}{ 0.8}{ 1.1} & \errtwo{ 0.5}{ 0.4}{ 0.3} &  78. \\
\noalign{\vspace*{0.5mm}} 
8 & D & \errtwo{18.}{ 4.}{ 2.} & \errtwo{ 0.37}{ 0.03}{ 0.05} & \errtwo{ 1.3}{ 0.3}{ 0.4} & \errtwo{  3.9}{  2.0}{  2.8} & \errtwo{-1.37}{ 0.17}{ 0.60} & \errtwo{22.}{ 4.}{ 5.} & \errtwo{ 1.1}{ 1.2}{ 0.6} & \errtwo{ 0.3}{ 0.4}{ 0.2} &  67. \\
\noalign{\vspace*{0.5mm}} 
8 & T & \errtwo{21.}{2.}{1.} & \errtwo{0.35}{0.03}{0.03} &
\errtwo{1.1}{0.2}{0.2} & \errtwo{5.7}{0.8}{1.0} & \errtwo{-1.33}
{ 0.05}{0.07} & \errtwo{20.}{ 2.}{ 1.} & \errtwo{ 2.3}{ 0.3}{ 0.3} &
\errtwo{0.7}{0.1}{0.1} &  74. \\
\noalign{\vspace*{0.5mm}} 
\hline
\end{tabular} } \hfill
\vspace*{3mm}  
%


%% file: tcoh.tex
\hfill\hbox{
\begin{tabular}{llllll}
\hline
\hline
\noalign{\vspace*{0.5mm}} 
Obs. & $\Delta \phi_{l1}$ & $\Delta \phi_{pl}$ & $\Delta \phi_{l2}$ & $\chi^2$/DoF & $\chi^2_{\rm red}$ \\ 
& (rad) & (rad) & (rad) \\
\noalign{\vspace*{0.5mm}} 
\hline
\noalign{\vspace*{0.5mm}} 
1 & 0.13 & 0.06 & 0.49 & 301.5/119 & 2.53 \\
\noalign{\vspace*{0.5mm}} 
2 & 0.18 & 0.05 & 0.26 & 283.4/119 & 2.38 \\
\noalign{\vspace*{0.5mm}} 
3 & 0.13 & 0.04 & 0.28 & 269.9/119 & 2.27 \\ 
\noalign{\vspace*{0.5mm}} 
4 & 0.15 & 0.06 & 0.66 & 195.8/119 & 1.65 \\
\noalign{\vspace*{0.5mm}} 
5 & 0.06 & 0.04 & 0.25 & 132.4/113 & 1.17 \\
\noalign{\vspace*{0.5mm}} 
6 & 0.16 & 0.05 & 0.38 & 217.4/119 & 1.82 \\
\noalign{\vspace*{0.5mm}}
7 & 0.11 & 0.04 & 0.27 & 309.9/119 & 2.60 \\
\noalign{\vspace*{0.5mm}} 
8 & 0.16 & 0.05 & 0.41 & 198.2/119 & 1.67 \\
\noalign{\vspace*{0.5mm}} 
\hline

\end{tabular} } \hfill
\vspace*{3mm}  
%


%% file: gxpap_time.bbl
\begin{thebibliography}{}

\bibitem[\protect\astroncite{Belloni \& Hasinger}{1990a}]{belloni:90a}
Belloni, T., \& Hasinger, G.,  1990a, A\&A, 227, L33

\bibitem[\protect\astroncite{Belloni \& Hasinger}{1990b}]{belloni:90b}
Belloni, T., \& Hasinger, G.,  1990b, A\&A, 230, 230

\bibitem[\protect\astroncite{Belloni et~al.}{1997}]{belloni:97a}
Belloni, T., {van der Klis}, M., Lewin, W. H.~G., {van Paradijs}, J., Dotani,
  T., Mitsuda, K., \& Miyamoto, S.,  1997, A\&A, 322, 857

\bibitem[\protect\astroncite{Bendat \& Piersol}{1986}]{bendat}
Bendat, J., \& Piersol, A.,  1986,
\newblock Random Data: Analysis and Measurement Procedures,
\newblock  (New York: Wiley)

\bibitem[\protect\astroncite{Corbel et~al.}{1997}]{corbel:98a}
Corbel, S., Fender, R.~P., Durouchoux, P., Sood, R.~K., Tzioumis, A.~K.,
  Spencer, R.~E., \& Campbell-Wilson, D.,  1997,
\newblock in Proc.\ 4th {C}ompton {S}ymposium, ed. C.~D. Dermer, M.~S.
  Strickman, J.~D. Kurfess,  (Woodbury: AIP),  937

\bibitem[\protect\astroncite{Cowley et~al.}{1991}]{cowley:91a}
Cowley, A.~P., et~al., 1991, ApJ, 381, 526

\bibitem[\protect\astroncite{Crary et~al.}{1998}]{crary:98a}
Crary, D.~J., Finger, M.~H., {van der Hooft}, C. K.~F., {van Paradijs}, J.,
  {van der Klis}, M., \& Lewin, W. H.~G.,  1998, ApJ,
\newblock submitted

\bibitem[\protect\astroncite{Cui et~al.}{1997}]{cui:97b}
Cui, W., Zhang, S.~N., Focke, W., \& Swank, J.~H.,  1997, ApJ, 484, 383

\bibitem[\protect\astroncite{Davies}{1990}]{davies:90a}
Davies, S.~R.,  1990, MNRAS, 244, 93

\bibitem[\protect\astroncite{Dove et~al.}{1997}]{dove:97b}
Dove, J.~B., Wilms, J., Maisack, M.~G., \& Begelman, M.~C.,  1997, ApJ, 487,
  759

\bibitem[\protect\astroncite{Esin, {McClintock} \& Narayan}{1997}]{esin:97c}
Esin, A.~A., {McClintock}, J.~E., \& Narayan, R.,  1997, ApJ, 489, 865

\bibitem[\protect\astroncite{Fender et~al.}{1997}]{fender:97a}
Fender, R.~P., Spencer, R.~E., Newell, S.~J., \& Tzioumis, A.~K.,  1997, MNRAS,
  286, L29

\bibitem[\protect\astroncite{Grebenev et~al.}{1991}]{grebenev:91a}
Grebenev, S.~A., Syunyaev, R., Pavlinsky, M.~N., \& Dekhanov, I.~A.,  1991,
  Sov. Astron. Lett., 17, 413

\bibitem[\protect\astroncite{Hannikainnen et~al.}{1998}]{hanni:98a}
Hannikainnen, D.~C., Hunstead, R.~W., Campbell-Wilson, D., \& Sood, R.~K.,
  1998, A\&A,
\newblock in press

\bibitem[\protect\astroncite{Horne \& Baliunas}{1986}]{horne:86a}
Horne, J.~H., \& Baliunas, S.~L.,  1986, ApJ, 302, 757

\bibitem[\protect\astroncite{Jahoda et~al.}{1996}]{jahoda:96b}
Jahoda, K., Swank, J.~H., Giles, A.~B., Stark, M.~J., Strohmayer, T., Zhang,
  W., \& Morgan, E.~H.,  1996,
\newblock in {EUV}, X-Ray, and Gamma-Ray Instrumentation for Astronomy {VII},
  ed. O.~H. Siegmund,  (Bellingham, WA: SPIE), 59

\bibitem[\protect\astroncite{Kazanas, Hua \& Titarchuk}{1997}]{kazanas:97a}
Kazanas, D., Hua, X.-M., \& Titarchuk, L.,  1997, ApJ, 480, 280

\bibitem[\protect\astroncite{Kemp et~al.}{1983}]{kemp:83a}
Kemp, J.~C., et~al., 1983, ApJ, 271, L65

\bibitem[\protect\astroncite{Leahy et~al.}{1983}]{leahy:83a}
Leahy, D.~A., Darbro, W., Elsner, R.~F., Weisskopf, M.~C., Sutherland, P.~G.,
  Kahn, S., \& Grindlay, J.,  1983, ApJ, 266, 160

\bibitem[\protect\astroncite{Levine et~al.}{1996}]{levine:96a}
Levine, A.~M., Bradt, H., Cui, W., Jernigan, J.~G., Morgan, E.~H., Remillard,
  R., Shirey, R.~E., \& Smith, D.~A.,  1996, ApJ, 469, L33

\bibitem[\protect\astroncite{Lochner \& Remillard}{1997}]{lochner:97a}
Lochner, J., \& Remillard, R.,  1997,
\newblock {ASM} Data Products Guide, Version Dated August 27, 1997,
  http://heasarc.gsfc.nasa.gov/docs/xte/asm\_products\_guide.html

\bibitem[\protect\astroncite{Lochner, Swank \& Szymkowiak}{1991}]{lochner:91a}
Lochner, J.~C., Swank, J.~H., \& Szymkowiak, A.~E.,  1991, ApJ, 376, 295

\bibitem[\protect\astroncite{Lomb}{1976}]{lomb:76a}
Lomb, N.~R.,  1976, Ap\&SS, 39, 447

\bibitem[\protect\astroncite{Maloney, Begelman \& Nowak}{1998}]{maloney:98a}
Maloney, P.~R., Begelman, M., \& Nowak, M.~A.,  1998, ApJ,
\newblock in press

\bibitem[\protect\astroncite{Maloney \& Begelman}{1997}]{maloney:97a}
Maloney, P.~R., \& Begelman, M.~C.,  1997, ApJ, 491, L43

\bibitem[\protect\astroncite{Maloney, Begelman \& Pringle}{1996}]{maloney:96a}
Maloney, P.~R., Begelman, M.~C., \& Pringle, J.~E.,  1996, ApJ, 472, 582

\bibitem[\protect\astroncite{M\'endez \& {van der Klis}}{1997}]{mendez:97a}
M\'endez, M., \& {van der Klis}, M.,  1997, ApJ, 479, 926

\bibitem[\protect\astroncite{Miyamoto et~al.}{1991}]{miyamoto:91a}
Miyamoto, S., Kimura, K., Kitamoto, S., Dotani, T., \& Ebisawa, K.,  1991, ApJ,
  383, 784

\bibitem[\protect\astroncite{Miyamoto \& Kitamoto}{1989}]{miyamoto:89a}
Miyamoto, S., \& Kitamoto, S.,  1989, Nature, 342, 773

\bibitem[\protect\astroncite{Miyamoto et~al.}{1992}]{miyamoto:92a}
Miyamoto, S., Kitamoto, S., Iga, S., Negoro, H., \& Terada, K.,  1992, ApJ,
  391, L21

\bibitem[\protect\astroncite{Narayan}{1996}]{narayan:96e}
Narayan, R.,  1996, ApJ, 462, 136

\bibitem[\protect\astroncite{Nowak}{1994}]{nowak:94a}
Nowak, M.~A.,  1994, ApJ, 422, 688,
\newblock (N94)

\bibitem[\protect\astroncite{Nowak et~al.}{1997}]{nowak:97a}
Nowak, M.~A., Vaughan, B.~A., Dove, J., \& Wilms, J.,  1997,
\newblock in Accretion Phenomena and Related Outflows, ed. D. Wickramasinghe,
  L. Ferrario, G.~V. Bicknell,  (San Francisco: Astron.\ Soc.\ Pacific),  366

\bibitem[\protect\astroncite{Nowak et~al.}{1998a}]{nowak:98a}
Nowak, M.~A., Vaughan, B.~A., Wilms, J., Dove, J., \& Begelman, M.~C.,  1998a,
  ApJ,
\newblock in press ~(paper II)

\bibitem[\protect\astroncite{Nowak et~al.}{1998b}]{nowak:98b}
Nowak, M.~A., Wilms, J., Vaughan, B.~A., Dove, J., \& Begelman, M.~C.,  1998b,
  ApJ,
\newblock in press

\bibitem[\protect\astroncite{Poutanen \& Fabian}{1998}]{poutanen:98a}
Poutanen, J., \& Fabian, A.~C.,  1998, MNRAS,
\newblock submitted (PF)

\bibitem[\protect\astroncite{Priedhorsky et~al.}{1979}]{priedhorsky:79a}
Priedhorsky, W.~C., Garmire, G.~P., Rothschild, R., Boldt, E., Serlemitsos, P.,
  \& Holt, S.,  1979, ApJ, 233, 350

\bibitem[\protect\astroncite{Priedhorsky, Terrell \&
  Holt}{1983}]{priedhorsky:83a}
Priedhorsky, W.~C., Terrell, J., \& Holt, S.~S.,  1983, ApJ, 270, 233

\bibitem[\protect\astroncite{Pringle}{1996}]{pringle:96a}
Pringle, J.~E.,  1996, MNRAS, 281, 357

\bibitem[\protect\astroncite{Remillard \& Levine}{1997}]{remillard:97a}
Remillard, R.~A., \& Levine, A.~M.,  1997,
\newblock in All-Sky {X}-Ray Observations in the Next Decade, ed. N. Matsuoka,
  N. Kawai,  (Tokyo: Riken), ~29

\bibitem[\protect\astroncite{Scargle}{1982}]{scargle:82a}
Scargle, J.~D.,  1982, ApJ, 263, 835

\bibitem[\protect\astroncite{Schwarzenberg-Czerny}{1989}]{schwarz:89a}
Schwarzenberg-Czerny, A.,  1989, MNRAS, 241, 153

\bibitem[\protect\astroncite{Sutherland, Weisskopf \&
  Kahn}{1978}]{sutherland:78a}
Sutherland, P.~G., Weisskopf, M.~C., \& Kahn, S.~M.,  1978, ApJ, 219, 1029

\bibitem[\protect\astroncite{Terrell}{1972}]{terrell:72a}
Terrell, N.~J.,  1972, ApJ, 174, L35

\bibitem[\protect\astroncite{{van der Klis}}{1989}]{vanderklis:89b}
{van der Klis}, M.,  1989,
\newblock in Timing Neutron Stars, ed. H. {\"Ogelman}, E.~P.~J. {van den
  Heuvel},  (Dordrecht: Kluwer), 27

\bibitem[\protect\astroncite{Vaughan \& Nowak}{1997}]{vaughan:97a}
Vaughan, B.~A., \& Nowak, M.~A.,  1997, ApJ, 474, L43

\bibitem[\protect\astroncite{Wilms et~al.}{1998a}]{wilms:98b}
Wilms, J., Nowak, M.~A., Dove, J.~B., Fender, R.~P., \& {D}i {M}atteo, T.,
  1998a, ApJ,
\newblock submitted (paper I)

\bibitem[\protect\astroncite{Wilms et~al.}{1998b}]{wilms:98a}
Wilms, J., Nowak, M.~A., Dove, J.~B., \& Heindl, B.,  1998b, ApJ,
\newblock in preparation

\bibitem[\protect\astroncite{Zdziarski et~al.}{1998}]{zdziarski:98a}
Zdziarski, A.~A., Poutanen, J., Miko\l{}ajewska, J., Gierli\'nski, M., Ebisawa,
  K., \& Johnson, W.~N.,  1998, MNRAS,
\newblock in press

\bibitem[\protect\astroncite{Zhang \& Jahoda}{1996}]{zhangw:96a}
Zhang, W., \& Jahoda, K.,  1996,
\newblock Deadtime Effects in the {PCA},
\newblock Technical report,  (Greenbelt: NASA GSFC),
\newblock version dated 1996 September 26

\bibitem[\protect\astroncite{Zhang et~al.}{1995}]{zhangw:95a}
Zhang, W., Jahoda, K., Swank, J.~H., Morgan, E.~H., \& Giles, A.~B.,  1995,
  ApJ, 449, 930

\end{thebibliography}
